\documentclass[twocolumn]{aastex631}
\usepackage{graphics,epsf}
\usepackage{amsmath}                
\usepackage{amsfonts}               
\usepackage{amssymb}                
\usepackage{epsfig}                 
\usepackage{appendix}
\usepackage{graphicx}
\usepackage{float}
\usepackage{color}
\usepackage{multirow}
\usepackage{colortbl}
\usepackage[para,online,flushleft]{threeparttable}
\usepackage{url}

\hypersetup{citecolor=blue, 
            linkcolor=red, 
            menucolor=blue, 
            urlcolor=blue}  
            
\usepackage{hyperref}

\newcommand{\m}{{~\rm m}}

\newcommand{\s}{{~\rm s}}
\newcommand{\h}{{~\rm h}}

\newcommand{\g}{{~\rm g}}
\newcommand{\G}{{~\rm G}}
\newcommand{\K}{{~\rm K}}

\newcommand{\yr}{{~\rm yr}}

\begin{document}

\title{Wobbling jets in common envelope evolution}


\author{Noam Dori}
\affiliation{Department of Physics, Technion, Haifa, 3200003, Israel; soker@physics.technion.ac.il; ealeal44@technion.ac.il}

\author{Ealeal Bear}
\affiliation{Department of Physics, Technion, Haifa, 3200003, Israel; soker@physics.technion.ac.il; ealeal44@technion.ac.il}

\author[0000-0003-0375-8987]{Noam Soker}
\affiliation{Department of Physics, Technion, Haifa, 3200003, Israel; soker@physics.technion.ac.il; ealeal44@technion.ac.il}

\begin{abstract}
We find that the convective motion in the envelopes of red supergiant (RSG) stars supplies a non-negligible stochastic angular momentum to the mass that a secondary star accretes in a common envelope evolution (CEE), such that jets that the secondary star launches wobble. The orbital motion of the secondary star in a CEE and the density gradient in the envelope impose a non-zero angular momentum to the accreted mass with a constant direction parallel to the orbital angular momentum. From one-dimensional stellar evolution simulations with the numerical code \textsc{mesa} we find that the stochastic convection motion in the envelope of RSG stars adds a stochastic angular momentum component with an amplitude that is about 0.1-1 times that of the constant component due to the orbital motion. We mimic a CEE of the RSG  star by removing envelope mass at a high rate and by depositing energy into its envelope. The stochastic angular momentum implies that the accretion disk around the secondary star (which we do not simulate), and therefore the jets that it launches, wobble with angles of up to tens of degrees with respect to the orbital angular momentum axis. This wobbling makes it harder for jets to break out from the envelope and can shape small bubbles in the ejecta that compress filaments that appear as arcs in the ejected nebula, i.e., in planetary nebulae when the giant is an asymptotic giant branch star. 
\end{abstract}


\section{Introduction}
\label{sec:intro}

 When one of the stars in a sufficiently close binary system evolves into a giant it might engulf its more compact companion. The two stars spirals around each other in a common envelope (e.g., \citealt{Paczynski1976}). Gravitational drag forces during this common envelope evolution (CEE) transfer orbital energy and angular momentum to the common envelope and the core of the giant and the compact companion spirals-in toward each other (for reviews see, e.g., \citealt{Ivanovaetal2013, DeMarcoIzzard2017, RoepkeDeMarco2023}). 
The engulfment might take place already in the sub-giant phase of the expanding star, along the red giant branch (RGB) or during the asymptotic giant branch (AGB) phase for low mass stars, or during the red supergiant (RSG) phase of massive stars. The compact companion might be any object from a planet to a black hole (BH).
The release of orbital energy ejects part or the entire envelope. The final outcome is either the merger of the compact companion with the core of the giant star, or a surviving binary system of the core and the companion.  

 One of the open questions in the CEE is whether there is another energy source to eject the envelope beside the orbital energy. One of the suggestions is that in some cases jets that the compact companion launches as it accretes mass from the common envelope facilitate envelope removal (for a review see \citealt{Soker2016Rev}). In this study we explore one aspect of this process.  

Compact objects that spiral-in inside the extended envelope of giant stars interact with an inhomogeneous envelope because of the steep density gradient and because of the orbital motion (rather than a linear motion). If the compact object accretes mass from the envelope during this CEE, the inhomogeneous interaction results in a net angular momentum of the accreted gas. If the radius of the compact object is sufficiently small the accreted mass forms an accretion disk around the compact object, or an accretion belt in cases where the specific angular momentum of the accreted mass is somewhat lower than the minimum value to form an accretion disk. The accretion disk, and even an accretion belt \citep{SchreierSoker2016, Soker2023BeltMS}, might launch jets inside the common envelope. 

Accretion disks in CEE are most likely to form around  neutron star (NS) companions (e.g., \citealt{ArmitageLivio2000, Chevalier2012}) and around BHs (e.g., \citealt{Soker2022Rev} for a review on jets launched by a NS/BH in CEE). Planetary nebulae that are shaped by jets indicate that main sequence companions can also launch jets during CEE (e.g., \citealt{BlackmanLucchini2014};  \citealt{Soker2016Rev} for a review), or possibly during a grazing envelope evolution (GEE; e.g., \citealt{Shiberetal2017}). Both the density gradient and the orbital motion possess a mirror symmetry about the orbital plane. Therefore, both effects act to have the net angular momentum axis perpendicular to the orbital plane. If the accretion disk launches jets the jets' axis will be perpendicular to the orbital plane. 

Numerical simulations are unable yet to include all processes,  accretion, jet-launching and the full CEE (\citealt{RoepkeDeMarco2023} for a recent review). Therefore, most three dimensional CEE studies do not include jets at all (e.g., a small fraction of existing studies include, \citealt{Passyetal2012, RickerTaam2012, Nandezetal2014, Staffetal2016MN, Kuruwitaetal2016, Ohlmannetal2016a,  Iaconietal2017b, Chamandyetal2018, LawSmithetal2020, Zouetal2020, GlanzPerets2021a, GlanzPerets2021b, GonzalezBolivar2022, Lauetal2022a, Lauetal2022b, Chamandyetal2023}). The studies that do include jets in CEE cannot follow the accretion process, or if they do, they are limited to a very short time of evolution (e.g., \citealt{MorenoMendezetal2017, ShiberSoker2018, LopezCamaraetal2019, Schreieretal2019inclined, Shiberetal2019, LopezCamaraetal2020MN, LopezCamaraetal2022MS, Zouetal2022, Schreieretal2023, Hilleletal2023}). 
In this study we do not refer to the launching of two oppositely collimated outflows (jets) by the distorted common envelope at the final phases of the CEE (\citealt{Soker1992, Zouetal2020, Morenoetal2022, Ondratscheketal2022}), but only to jets that the companion launches as it accretes mass in CEE.  

Other processes beside the envelope density gradient and orbital motion make the launching of jets and their interaction with the envelope more complicated. Firstly, the jets have a negative feedback on the accretion process as they remove gas from the accreting companion vicinity  (e.g., \citealt{Soker2016Rev, Gricheneretal2021, Hilleletal2022FB}). 
Secondly, the envelope of giant stars have vigorous convection which introduce random velocity fluctuations into the accreted mass. The random velocity component introduces random component of angular momentum to the accreted mass and can cause the jets' axis to change direction erratically, e.g., wobbling jets. The random component of the accreted angular momentum because of envelope convection is the subject of this study. 
 
We note that the vigorous convection in giant stars plays roles in other processes, such as mass loss (e.g., \citealt{LaiCao2022, FreytagHofner2023}) and tidal interaction that can cause orbits of planets to become slightly eccentric (e.g., \citealt{Lanzaetal2023}).
 In the collapse of an RSG to a BH the envelope convective motion can lead to the formation of intermittent accretion disk around the newly born BH that launches jets in randomly different directions (e.g., \citealt{AntoniQuataert2022}), namely, jittering jets. As for CEE, convection can be very efficient in transferring to the photosphere the energy that the inspiral process releases, where the energy is radiated away (e.g., \citealt{WilsonNordhaus2019}). This process reduces the efficiency of envelope removal and by that influences the outcome of the CEE (e.g., \citealt{WilsonNordhaus2020} for the final separation of double white dwarf systems). In massive stars the convective energy transport during CEE is less efficient than in low mass stars \citep{WilsonNordhaus2022}.  

 We start by describing the sources of the angular momentum of the mass that the compact companion accretes during the CEE (section \ref{sec:CEE1}). We then present one-dimensional simulation of stellar models to explore the  convection properties relevant to our study (section \ref{sec:Convective}). In section \ref{sec:CEE2} we use the results of section \ref{sec:Convective}  in the derivation of section \ref{sec:CEE1}  to estimate the degree of the stochastic angular momentum component relative to that due to the density gradient in the envelope.  We discuss and summarize our main results in section \ref{sec:Summary}.

\section{Angular momentum sources of accreted mass}
\label{sec:CEE1}
 In this section we describe the two angular momentum sources of the accreted mass, that due to the orbital motion in a non-homogeneous envelope and that due to the convective motion in the envelope. We will obtain numerical values in section \ref{sec:CEE2} where we will use the results of stellar models that we simulate in section \ref{sec:Convective}.    
\subsection{Fixed angular momentum from orbital motion}
\label{subsec:OrbitAM}
We deal with a CEE of a compact secondary star that spirals-in inside the envelope of a primary giant star. The secondary star might be a main sequence star, a WD, a NS, or a BH. The giant star might be an RGB star, an AGB star, or an RSG star. We will scale equations for RSG stars, i.e., mainly $M_1 \simeq 10-30 M_\odot$, and a NS/BH companion or a main sequence companion, i.e., mainly $M_2 \simeq 1.4-10 M_\odot$. The orbital separation $a(t)$ might decrease, i.e., a spiralling-in motion, or be constant, like in the grazing envelope evolution (GEE). 
We consider the case where the secondary star of mass $M_2$ accretes mass from the envelope under the following assumptions \citep{Soker2004AM}. (1) The accretion process is a Bondi-Hoyle-Lyttleton (BHL) type accretion flow. The accretion rate might be lower than the BHL value (see below). (2) The velocity of the secondary star relative to the giant's envelope is the local Keplerian velocity of a test particle 
\begin{equation}
v_r = v_{\rm Kep}= \left[ \frac{G M_1(a)}{a} \right]^{1/2}, 
\label{eq:Vkep}
\end{equation}
where $M_1(a)$ is the giant mass inside radius $r=a$. (3) Because of the density gradient in the envelope and the circular (more or less) motion the gas that the secondary star accretes has a finite specific angular momentum $j_{\rm O}$ (subscript `O' stands for orbital motion). We take the value of $j_{\rm O}$ from the three-dimensional hydrodynamical simulations of \cite{Livioetal1986} who study the BHL accretion process in a rectangle numerical box and did not include neither the orbital motion of the secondary star nor the gravity of the giant. Like \cite{Soker2004AM} we assume that these results hold also for a CEE (or a GEE). 

\cite{Livioetal1986} simulate the accretion from a medium with a density gradient $\rho=\rho_0(1+y/H)$, where $y$ is perpendicular to the direction of the relative motion of the secondary star relative to the gas . 
For a motion through an envelope with a density profile of  
\begin{equation}
\rho_{\rm env} (r) \propto r^{-\beta}
\label{eq:EnvDensty}
\end{equation}
the density gradient at $r=a$ is $H=a/\beta$. For our unperturbed envelope $\beta \simeq 3$ (section \ref{sec:Convective}). 
\cite{Livioetal1986} find the specific angular momentum of the accreted gas to be 
\begin{equation}
j_{\rm O}=\frac{\eta}{4 H} 
\frac{ (2 G M_2)^2}{v^3_r} = \eta \beta \left[ \frac{M_2}{M_1(a)} \right]^{2} a v_{\rm Kep} 
\label{eq:jacc1}
\end{equation}
with a typical value of $\eta \simeq 0.25$ and where in the second equality we used equations (\ref{eq:Vkep}) and (\ref{eq:EnvDensty}). 
This angular momentum component of the accreted gas has a constant direction perpendicular to the orbital plane. 

We are interested in the launching of jets, for which we assume that the accretion is through an accretion disk, i.e., supported by the centrifugal force alone, or an accretion belt, where the specific angular momentum is somewhat lower than the minimum value to form an accretion disk. For an accretion disk to form the specific angular momentum of the accreted gas should be $\ga j_2 \equiv \sqrt{G M_2 R_2}$, where $R_2$ is the radius of the secondary star. With the help of the second equality in equation (\ref{eq:jacc1}) the condition to form an accretion disk due to the orbital motion alone is 
\begin{equation}
R_{\rm d2} \equiv \frac {j^2_{\rm O}}{GM_2}=\eta^2 \beta^2
\left[ \frac{M_2}{M_1(a)} \right]^{3} 
a \ga R_2 ,
\label{eq:EtaBeta1}
\end{equation}
where $R_{\rm d2}$ is defined by the first equality. 

\subsection{Stochastic angular momentum from convection}
\label{subsec:Stochastic}

In deriving $j_{\rm O}$ (equation \ref{eq:jacc1}) we assumed that the specific angular momentum of the accreted gas relative to the envelope is zero. However, because of the convective motion in the envelope it is not zero. We turn to estimate the additional specific angular momentum of the accreted gas because of the convective motion. 

Consider the presence of convective blobs with random velocities in the accreted gas. At each specific time about $N_{\rm bdisk}$ blobs contribute to the accretion disk. 
The typical velocity of a blob is the convective velocity $v_{\rm conv}$. As the blobs are accreted from a distance of up to about the BHL accretion radius $R_{\rm BHL}=2GM_2/v^2_r$, we take the typical impact parameter of a blobs to be $X_{\rm b} \simeq  R_{\rm BHL}/2$. For the number of blobs within the accretion radius we have 
$N_{\rm BHL} \simeq (R_{\rm BHL}/\lambda)^3$, where $\lambda$ is the mixing length.  
These blobs are accreted during a timescale that is the accretion time scale $\tau_{\rm BHL}\simeq R_{\rm BHL}/v_r$. However, the relevant number of blobs that determine the specific angular momentum of the accretion disk are the number of blobs that are accreted during a typical life time of an accretion disk, $\tau_{\rm disk}$. This number of blobs is then 
\begin{equation}
N_{\rm bdisk} \simeq N_{\rm BHL} \frac{\tau_{\rm disk}}{\tau_{\rm BHL}} 
\simeq \left( \frac{R_{\rm BHL}}{\lambda} \right)^3  \frac{\tau_{\rm disk} v_r}{R_{\rm BHL}}. 
\label{eq:Nbdisk} 
\end{equation}
If this equation gives $N_{\rm bdisk}<1$ then we must take $N_{\rm bdisk}=1$.
The direction of the specific angular momentum changes stochastically in all directions, and its magnitude is 
\begin{equation}
j_{\rm R} \simeq \frac{1}{\sqrt{N_{\rm bdisk}}} X_{\rm b} v_{\rm conv}  ,
\label{eq:jR1}
\end{equation}
where `R' stands for a random component. 

For the life time of the accretion disk we consider two alternatives. 
The first one is for a case of a large companion, namely a main sequence star. In this case we assume that the disk is very close to the surface and does not extend much. We estimate the life time of a parcel of gas in the accretion disk to be tens of times the orbital period on the surface of the secondary star 
\begin{equation}
\tau_{\rm disk,2} = \zeta \frac{ 2 \pi R_2^{3/2}}{(G M_2)^{1/2}}, 
\label{eq:taudisk2}
\end{equation}
 where this equation defines $\zeta$ to be ratio of the disk life-time to the Keplerian orbital period on the surface of the secondary star. 

Substituting equation (\ref{eq:Nbdisk}) in equation (\ref{eq:jR1}) with $\tau_{\rm disk}$ from equation (\ref{eq:taudisk2}) and with the other variables as above, we find the magnitude of the stochastic component of the accreted mass 
\begin{equation}
\begin{split}
j_{\rm R,2} \simeq  \frac{1}{\sqrt{8 \pi} \sqrt{\zeta}}
&   \left( \frac{\lambda}{a} \right)^{3/2}
\left( \frac{a}{R_2} \right)^{3/4}
\left[ \frac{M_2}{M_1(a)} \right]^{1/4} \\
\times & 
\left( \frac{v_{\rm conv}}{v_{\rm Kep}} \right)   a v_{\rm Kep}. 
\label{eq:jR2}
\end{split}
\end{equation}
For this case we find from equation (\ref{eq:Nbdisk})
\begin{equation}
\begin{split}
N_{\rm bdisk,2} & \simeq 8.3 
\left( \frac{\zeta}{100} \right) 
\left( \frac{a}{100R_2} \right)^{-3/2}
\\ \times &  
\left( \frac{\lambda}{0.3a} \right)^{-3}
\left[ \frac{M_2}{0.2M_1(a)} \right]^{3/2} .
\label{eq:Nbdisk2}
\end{split}
\end{equation}
As stated, if $N_{\rm bdisk,2}<1$ we take $N_{\rm bdisk,2} =1$ and apply this in equation (\ref{eq:jR2}). Namely, if $N_{\rm bdisk,2}<1$ we multiply the value of equation (\ref{eq:jR2}) by $\sqrt{N_{\rm bdisk,2}}$.  

In the second alternative the radius of the accretion disk is much larger than the radius of the mass-accreting compact star, mainly a NS or a BH. The radius of the disk is $R_{\rm d2}$ as given by equation (4). In this case the typical life time of the accretion disk is 
\begin{equation}
\begin{split}
\tau_{\rm disk,d} &= \zeta \frac{ 2 \pi R_{d2}^{3/2}}{(G M_2)^{1/2}} 
\\ &  =  
\zeta  
\eta^3 \beta^3
\left[ \frac{M_2}{M_1(a)} \right]^{4}
\frac{ 2 \pi a^{3/2}}{(G M_1)^{1/2}}. 
\label{eq:taudiskD}
\end{split}
\end{equation}
The last term is the Keplerian orbital period. In this case we find the stochastic specific angular momentum component to be 
\begin{equation}
\begin{split}
j_{\rm R,d} \simeq  \frac{\eta^{-3/2} \beta^{-3/2}}{\sqrt{8 \pi} \sqrt{\zeta}} 
&   \left( \frac{\lambda}{a} \right)^{3/2}
\left[ \frac{M_1(a)}{M_2} \right]^{2} \\
\times & 
\left( \frac{v_{\rm conv}}{v_{\rm Kep}} \right)   a v_{\rm Kep}. 
\label{eq:jRd}
\end{split}
\end{equation}
For this case we find from equation (\ref{eq:Nbdisk})
\begin{equation}
\begin{split}
N_{\rm bdisk,d} & \simeq 1.3 
\left( \frac{\zeta}{100} \right) 
\left( \frac{\beta}{3} \right)^{3}
\left( \frac{\eta}{0.2} \right)^{3} 
\\ \times &  
\left( \frac{\lambda}{0.3a} \right)^{-3}
\left[ \frac{M_2}{0.2M_1(a)} \right]^{6} .
\label{eq:NbdiskD}
\end{split}
\end{equation}
If $N_{\rm bdisk,d}<1$ we take $N_{\rm bdisk,d} =1$ and apply this in equation (\ref{eq:jRd}), i.e., we multiply the value of equation (\ref{eq:jRd}) by $\sqrt{N_{\rm bdisk,d}}$.
\subsection{The wobbling magnitude}
\label{subsec:Wobbling}

The total accreted angular momentum is 
\begin{equation}
\overrightarrow{j}_{\rm acc} = \overrightarrow j_{\rm O} + \overrightarrow{j}_{\rm R}. 
\label{eq:Jacc}
\end{equation}
The component $\overrightarrow{j}_{\rm R}$, which varies stochastically on a shorter timescale relative to the inspiral timescale, changes the magnitude and direction of $\overrightarrow {j}_{\rm O}$ that maintains a fixed direction and the magnitude changes on the longer timescale of the spiralling-in time. 
We find the ratio of the magnitudes of these two components from equations (\ref{eq:jacc1}) and (\ref{eq:jR2}) or (\ref{eq:jRd}). We scale this ratio by the typical quantities that we obtain in section \ref{sec:Convective}. For a disk on the surface of the secondary star, which might be appropriate for a main sequence companion we find   
\begin{equation}
\begin{split}
\frac{j_{\rm R,2}}{j_{\rm O}} & \simeq  0.29 
\left( \frac{\zeta}{100} \right)^{-1/2} 
\left( \frac{\beta}{3} \right)^{-1}
\left( \frac{\eta}{0.2} \right)^{-1} 
\left( \frac{\lambda}{0.3a} \right)^{3/2}
\\ \times &   
\left( \frac{a}{100 R_2} \right)^{3/4}
\left[ \frac{M_1(a)}{5M_2} \right]^{7/4} 
\left( \frac{v_{\rm conv}}{0.1 v_{\rm Kep}} \right)  . 
\label{eq:jRatio2}
\end{split}
\end{equation}
If $N_{\rm bdisk,2}<1$ we multiply the value of equation (\ref{eq:jRatio2}) by $\sqrt{N_{\rm bdisk,2}}$. We find that for the parameters we use here and for cases relevant to the CEE, namely with mass removal deposition $N_{\rm bdisk,2} > 1$. Only in the undisturbed model, before the onset of the CEE, and after energy deposition, is $N_{\rm bdisk,2}$ below unity, but still it is $N_{\rm bdisk,2} > 0.5$. After energy deposition, we have $N_{\rm bdisk,2} > 0.7$.

For the more general case where the radius of the accretion disk is according to equation (\ref{eq:EtaBeta1}) and the disk timescale is by equation (\ref{eq:taudiskD}) we find 
\begin{equation}
\begin{split}
\frac{j_{\rm R,d}}{j_{\rm O}} & \simeq  0.73 
\left( \frac{\zeta}{100} \right)^{-1/2} 
\left( \frac{\beta}{3} \right)^{-5/2}
\left( \frac{\eta}{0.2} \right)^{-5/2} 
\\ \times &  
\left( \frac{\lambda}{0.3a} \right)^{3/2}
\left[ \frac{M_1(a)}{5M_2} \right]^{4} 
\left( \frac{v_{\rm conv}}{0.1 v_{\rm Kep}} \right)  . 
\label{eq:jRatioD}
\end{split}
\end{equation}
If $N_{\rm bdisk,d}<1$ we multiply the value of equation (\ref{eq:jRatioD}) by $\sqrt{N_{\rm bdisk,d}}$.
We find that for the parameters we use here and for cases relevant to the CEE with mass removal and energy deposition $N_{\rm bdisk,d} > 0.5$. Before the onset of the CEE, i.e., in the undisturbed model we find that this values in some parts of the envelope is as low as $N_{\rm bdisk,d} \simeq 0.1$.

\section{Convective properties}
\label{sec:Convective}

 In this section we simulate stellar models that we will use in section \ref{sec:CEE2} to obtain numerical values from the expressions we derived in section \ref{sec:CEE1}. 

We used version 23.05.1-rc2 of the stellar evolution code Modules for Experiments in Stellar Astrophysics (\textsc{mesa}; \citealt{Paxtonetal2011, Paxtonetal2013, Paxtonetal2015, Paxtonetal2018, Paxtonetal2019}) in its single star mode. We evolve a stellar model with initial, zero age main sequence (ZAMS), mass of $M_1=15M_\odot$ and metallicity of $z=0.02$, to the RSG phase, based on the example of $\textit{20M\_pre\_ms\_to\_core\_collapse}$. All other parameters remain as in the default of \textsc{mesa}.\footnote{  
The default capabilities of \textsc{mesa}-single relay on the MESA EOS that is a blend of the OPAL \citep{RogersNayfonov2002}, SCVH
\citep{Saumonetal1995}, FreeEOS \citep{Irwin2004}, HELM \citep{TimmesSwesty2000}, PC \citep{PotekhinChabrier2010}, and Skye \citep{Jermynetal2021} EOSes. Radiative opacities are primarily from OPAL \citep{IglesiasRogers1993, IglesiasRogers1996}, with low-temperature data from \citet{Fergusonetal2005} and the high-temperature, Compton-scattering dominated regime by
\citet{Poutanen2017}.  Electron conduction opacities are from
\citet{Cassisietal2007} and \citet{Blouinetal2020}.
Nuclear reaction rates are from JINA REACLIB \citep{Cyburtetal2010}, NACRE \citep{Anguloetal1999} and additional tabulated weak reaction rates \citealt{Fulleretal1985, Odaetal1994, Langankeetal2000}.  Screening is included via the prescription of \citet{Chugunovetal2007}. Thermal neutrino loss rates are from \citealt{Itohetal1996}.}

We evolve the star up to a radius of $R_1=500R_\odot$, at which point the stellar mass is $M_1=14.7M_\odot$, the envelope mass is $M_{\rm env}=11.4M_\odot$ and the core mass is $M_{\rm core}=3.3M_\odot$. The stellar luminosity at this time is $L_1= 3\times10^4 L_\odot$ and helium burns in the core and hydrogen burns in a shell. 
At that evolutionary time we start to remove mass at very high rates of either $\dot{M}= 0.01$, $0.1$ or $1 M_\odot \yr^{-1}$. These high mass removal rates mimic a CEE where a companion removes mass as it spirals-in inside the envelope. Jets that the companion launches might play a role as well in the rapid mass removal process, even a dominant role (section \ref{sec:intro}). 

In what follows we set $t=0$ at the time when we end the rapid mass removal. 

We assume that jets deposit more energy to the remaining envelope in addition to the energy that was required to remove the envelope. We therefore deposit energy to the envelope after we removed the mass. Although the deposition of energy and the ejection of envelope mass occur simultaneously, we here study the effects of each of these processes, and for that deposit energy after we remove the mass. We do not reproduce CEE, but rather interested only in the properties of the convection, velocity and mixing length, that we use to our calculations in section \ref{sec:CEE2}. We use these for scaling and interested in particular in NS/BH companions that launch jets, and for that we present results for only one massive star model. 

As we do not follow the interaction of jets in a CEE, we have to build a prescription to deposit energy to the envelope. We proceed as follows. We inject energy at a power of $\dot E$ and distribute it in a fraction of the envelope with a constant power per unit mass. We use one of these powers, $\dot E=2\times 10^5L_\odot$, $4\times 10^5L_\odot$, $8\times 10^5L_\odot$, or $16\times 10^5L_\odot$.  This range of powers is on the lower side of what \cite{Hilleletal2022FB} take in their there-dimensional hydrodynamical simulations, because we assume that a large fraction of the jets' energy is taken away by the expelled mass. Namely, only some fraction of the original jets' energy is left in the still bound envelope.  We distribute this energy into a shell within the envelope. In some simulations we take this shell to be $0.07 R_1(t) < r < R_1(t)$ while in others  $0.25R_1(t) < r < R_1(t)$. 

We present the convective properties in a way that will serve us in our calculations in section \ref{sec:CEE2} where we use expressions from section \ref{sec:CEE1}. These quantities are the ratio of the convective velocity to the Keplerian velocity, $v_{\rm Kep}(r)=\sqrt{G M_1(r)/r}$, at the given radius, $v_{\rm conv}(r)/v_{\rm Kep}(r)$, and the mixing length to radius ratio $\lambda(r)/r$.  
   
\begin{figure} 
	\centering
\includegraphics[trim=3.2cm 9.4cm 4cm 9.9cm ,clip, scale=0.54]{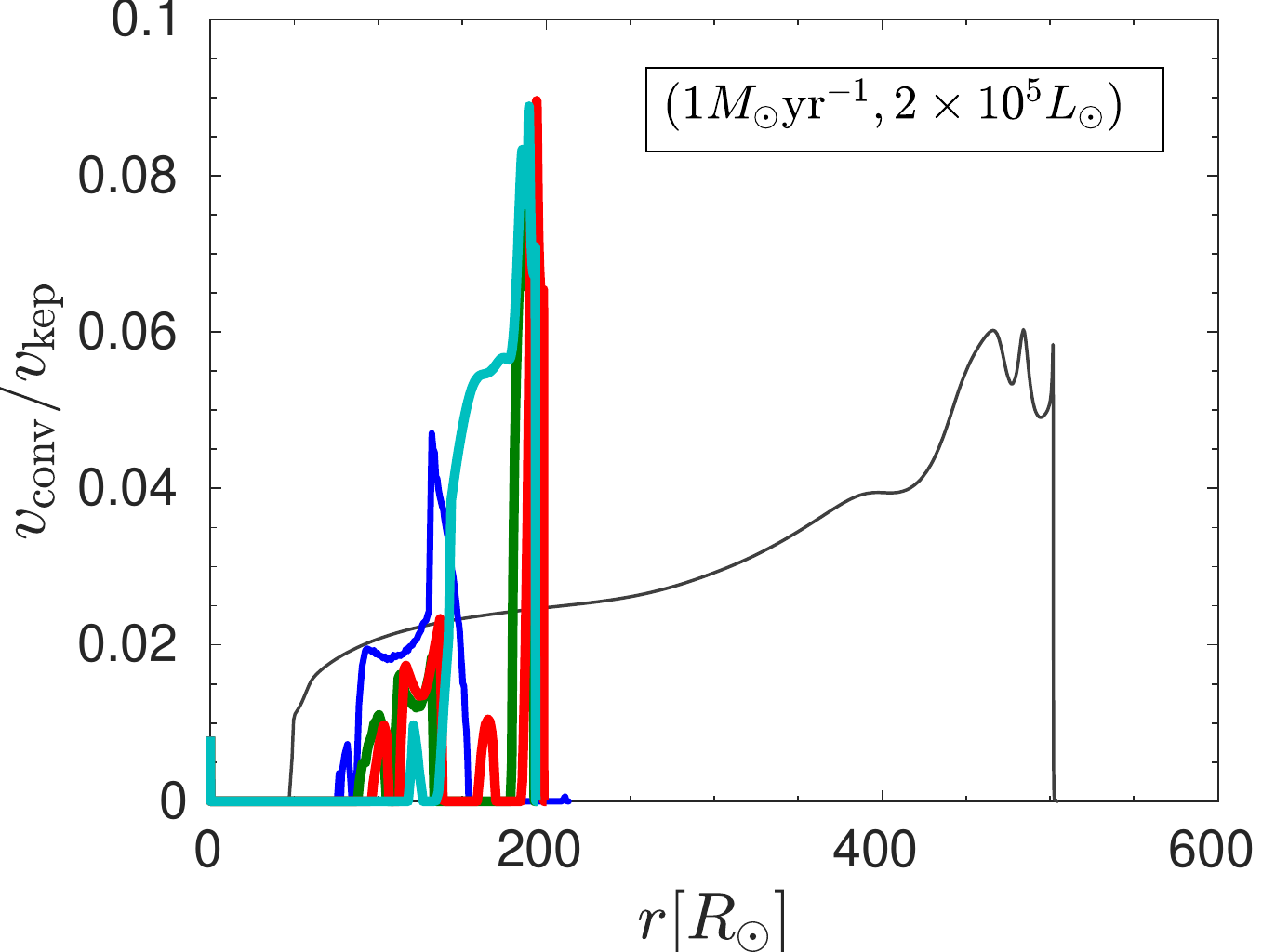}
\includegraphics[trim=3.2cm 9.3cm 4cm 9.9cm ,clip, scale=0.54]{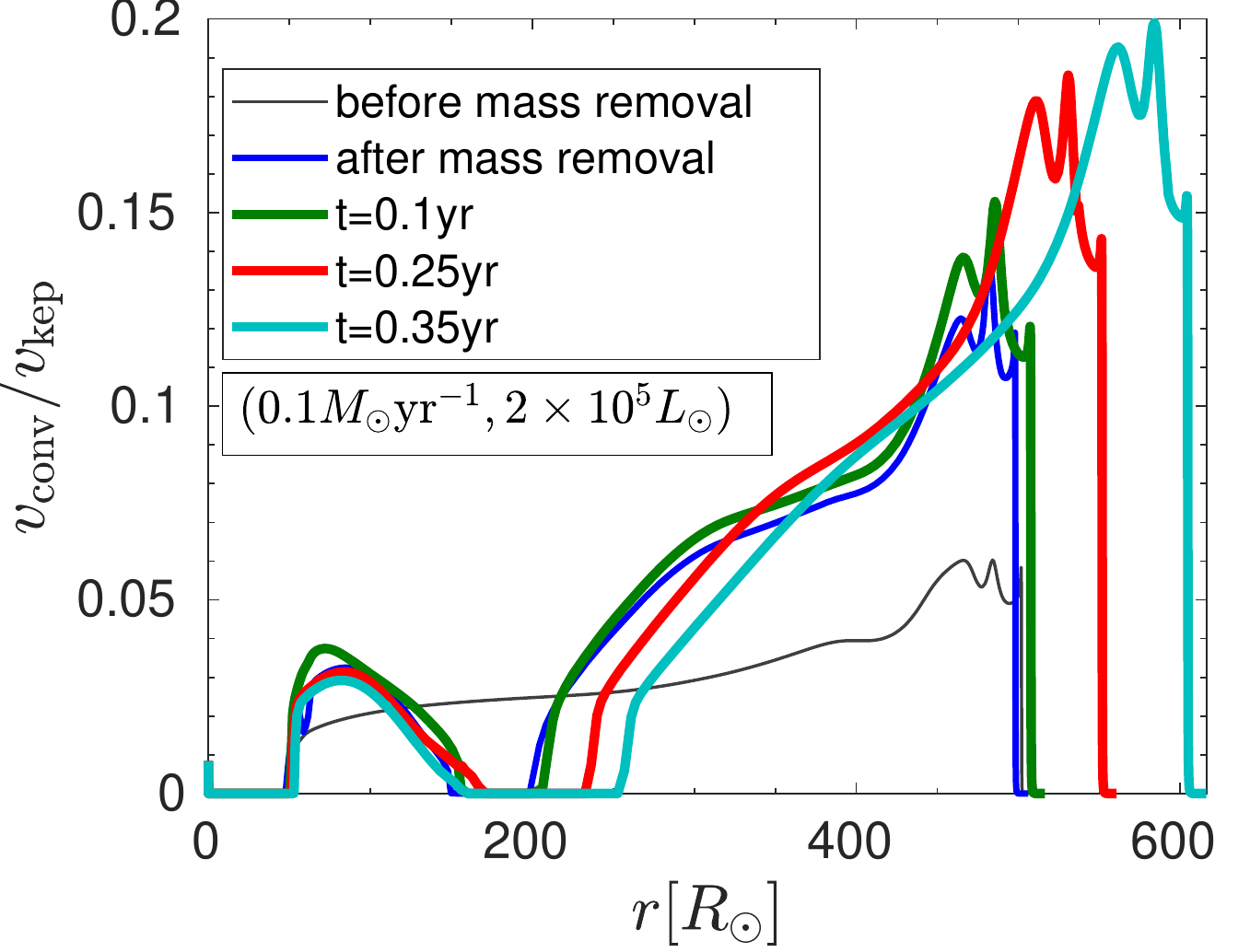}
\includegraphics[trim=3.2cm 9.4cm 4cm 9.9cm ,clip, scale=0.54]{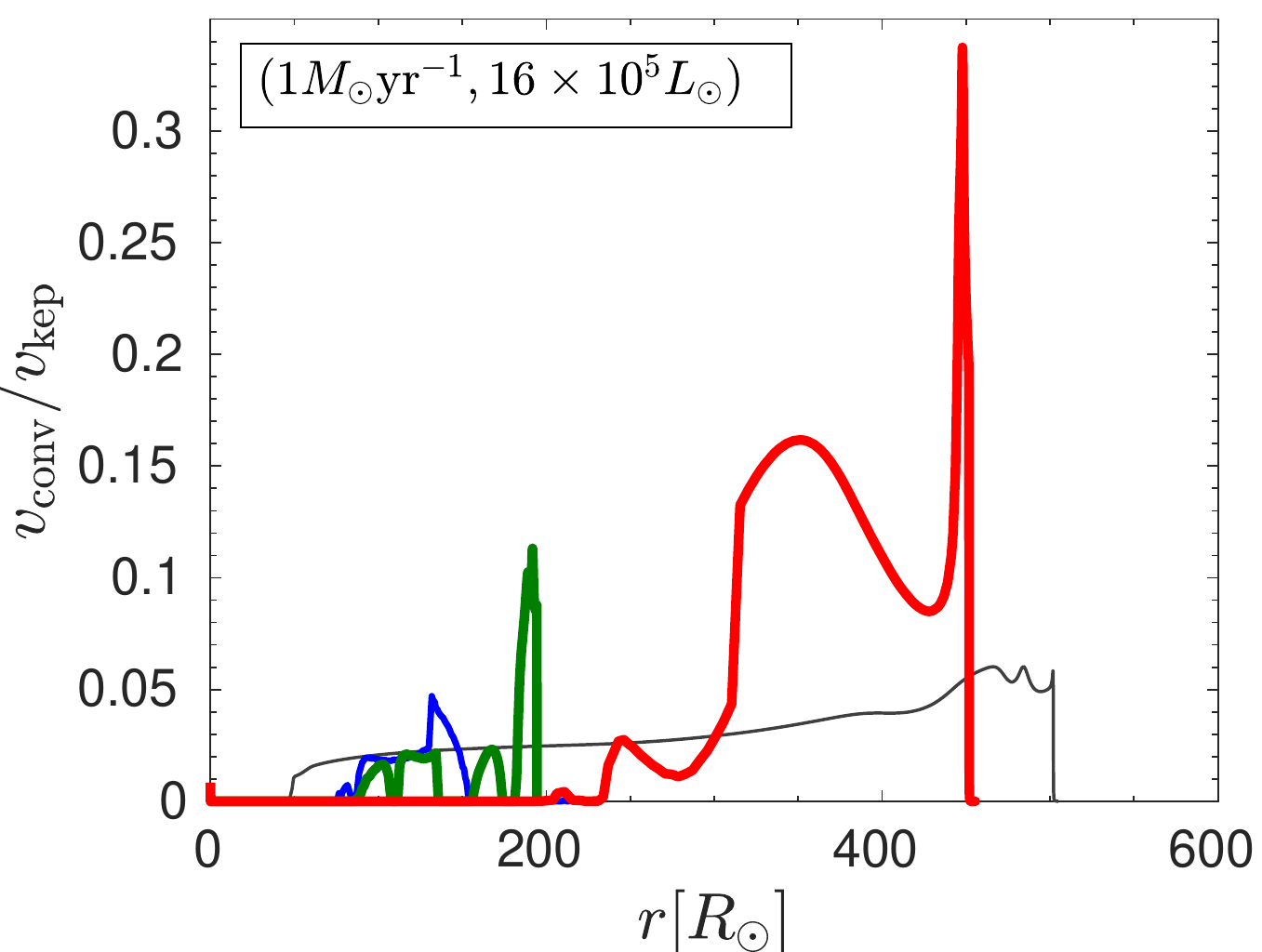}
\caption{The ratio of the convective velocity to the Keplerian velocity, $v_{\rm conv}(r)/v_{\rm Kep}(r)$, as a function of radius inside the envelope. The thin-black line in each panel depicts this quantity when we just start to remove mass at a high rate. The blue line represents the envelope after we removed an envelope mass of $\Delta M_{\rm env}=9.5 M_\odot$ to have $M_{\rm env} (0)=1.9 M_\odot$; we set $t=0$ when mass removal ends. At $t=0$ we start to inject energy from radius $R_{\rm in}(t)=0.07 R_1(t)$ to $R_1(t)$, where $R_1(t)$ is the stellar radius. The green, red, and cyan lines are at $t=0.1 \yr$,$t=0.25 \yr$, and $t=0.35 \yr$ respectively. The insets give the mass removal rate and the power of energy injection $(\dot M, \dot E_{\rm in})$. The last panel does not have the line at $t=0.35 \yr$ because the expansion at that time is on a shorter than the dynamical timescale, and the results unreliable. Note the different scaling of the axes in the different panels.}
\label{fig:Vconv_93p_2mf}
\end{figure}

\begin{figure}[t]
	\centering
\includegraphics[trim=3.2cm 9.4cm 3.5cm 9.9cm ,clip, scale=0.54]{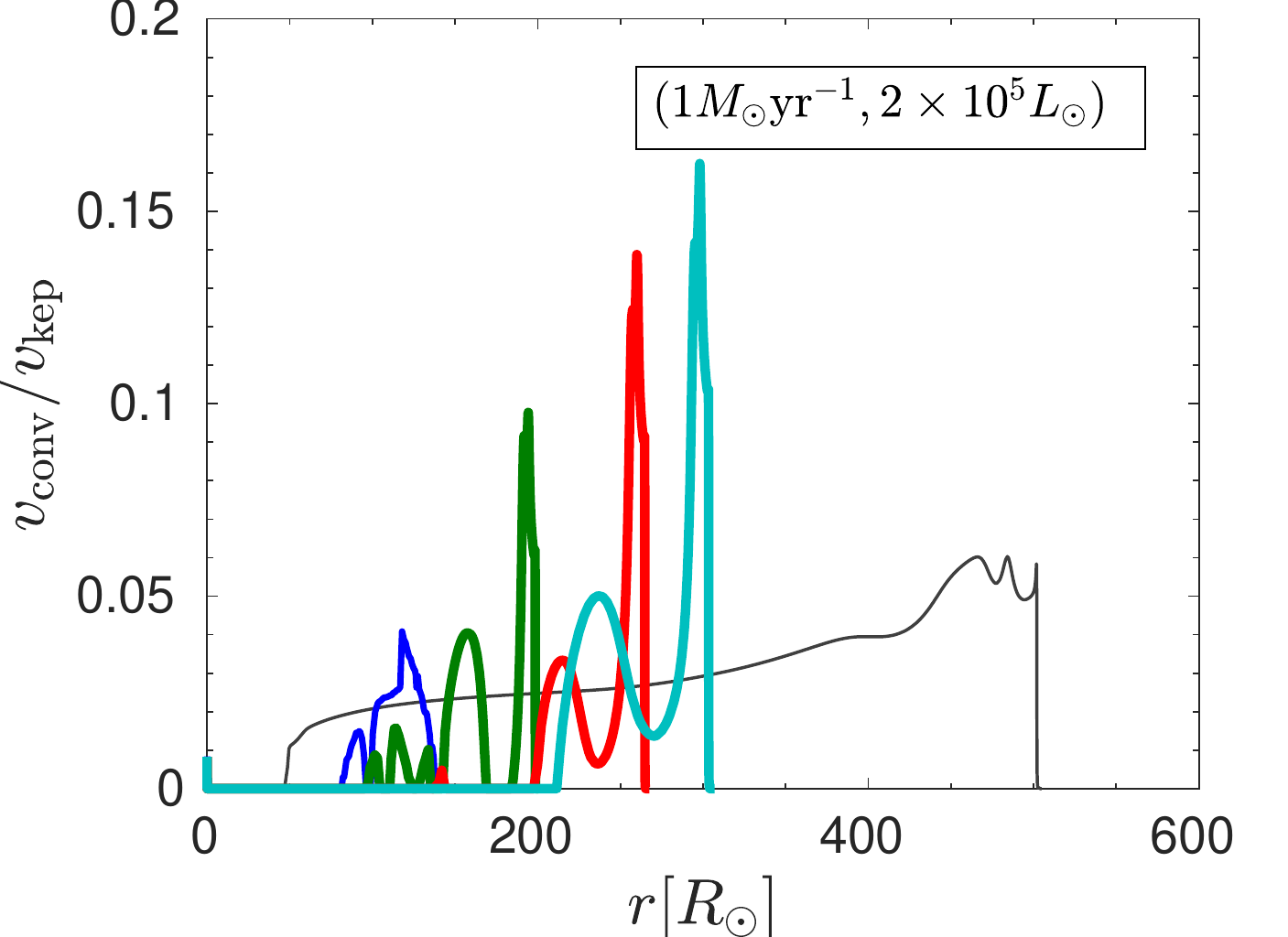}
\includegraphics[trim=3.2cm 9.3cm 3.5cm 9.9cm ,clip, scale=0.54]{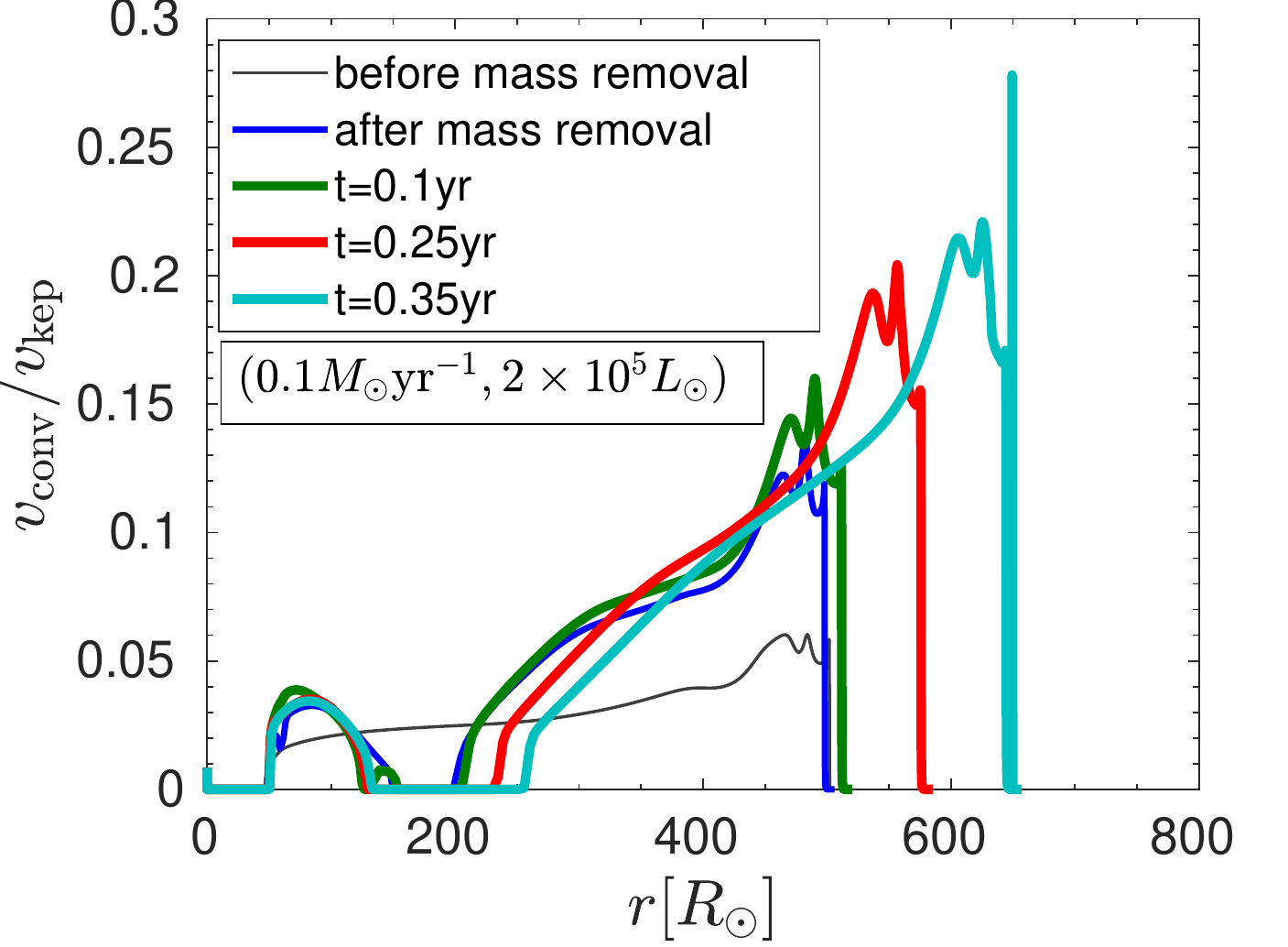}
\includegraphics[trim=3.2cm 9.4cm 3.5cm 9.9cm ,clip, scale=0.54]{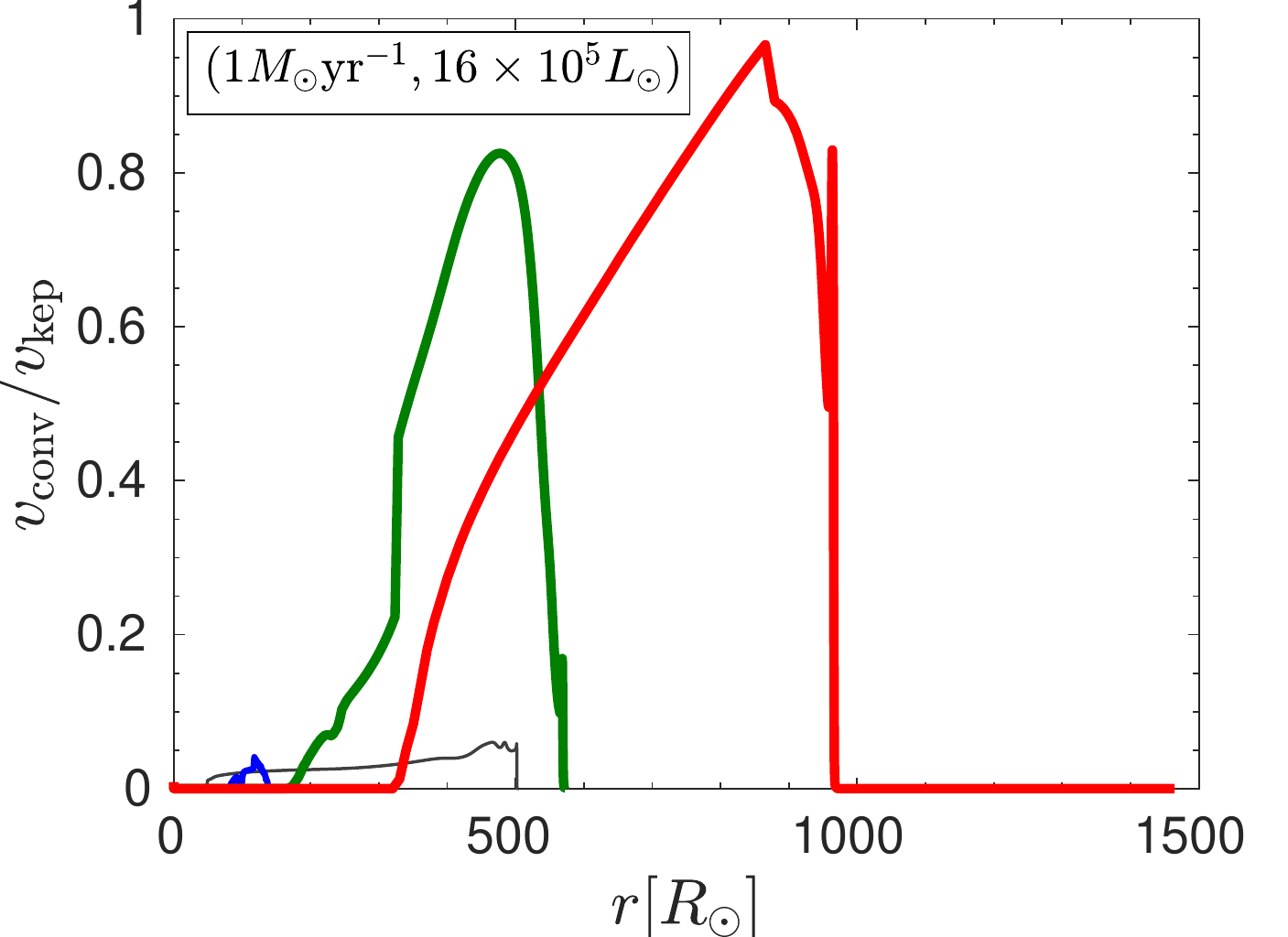}
\caption{Similar to Fig. \ref{fig:Vconv_93p_2mf}, but the energy injection is to a shell $R_{\rm in}(t)=0.25 R_1(t) < r < R_1(t)$ (rather than into a shell $R_{\rm in}(t)=0.07 R_1(t) < r < R_1(t)$).
}
\label{fig:Vconv_75p_2mf}
\end{figure}

\begin{figure}[t]
	\centering
\includegraphics[trim=3.2cm 9.4cm 4cm 9.9cm ,clip, scale=0.54]{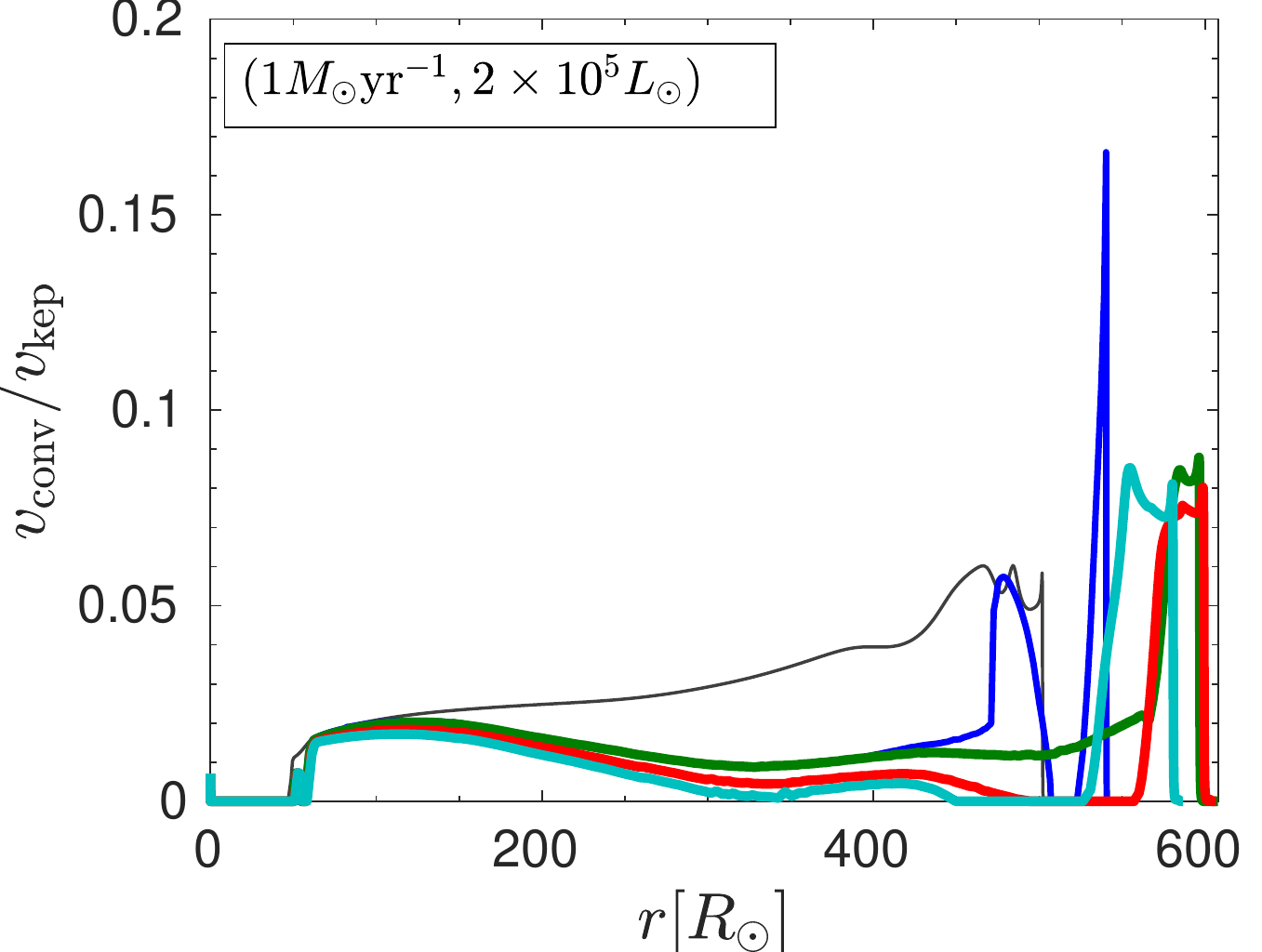}
\includegraphics[trim=3.2cm 9.3cm 4cm 9.9cm ,clip, scale=0.54]{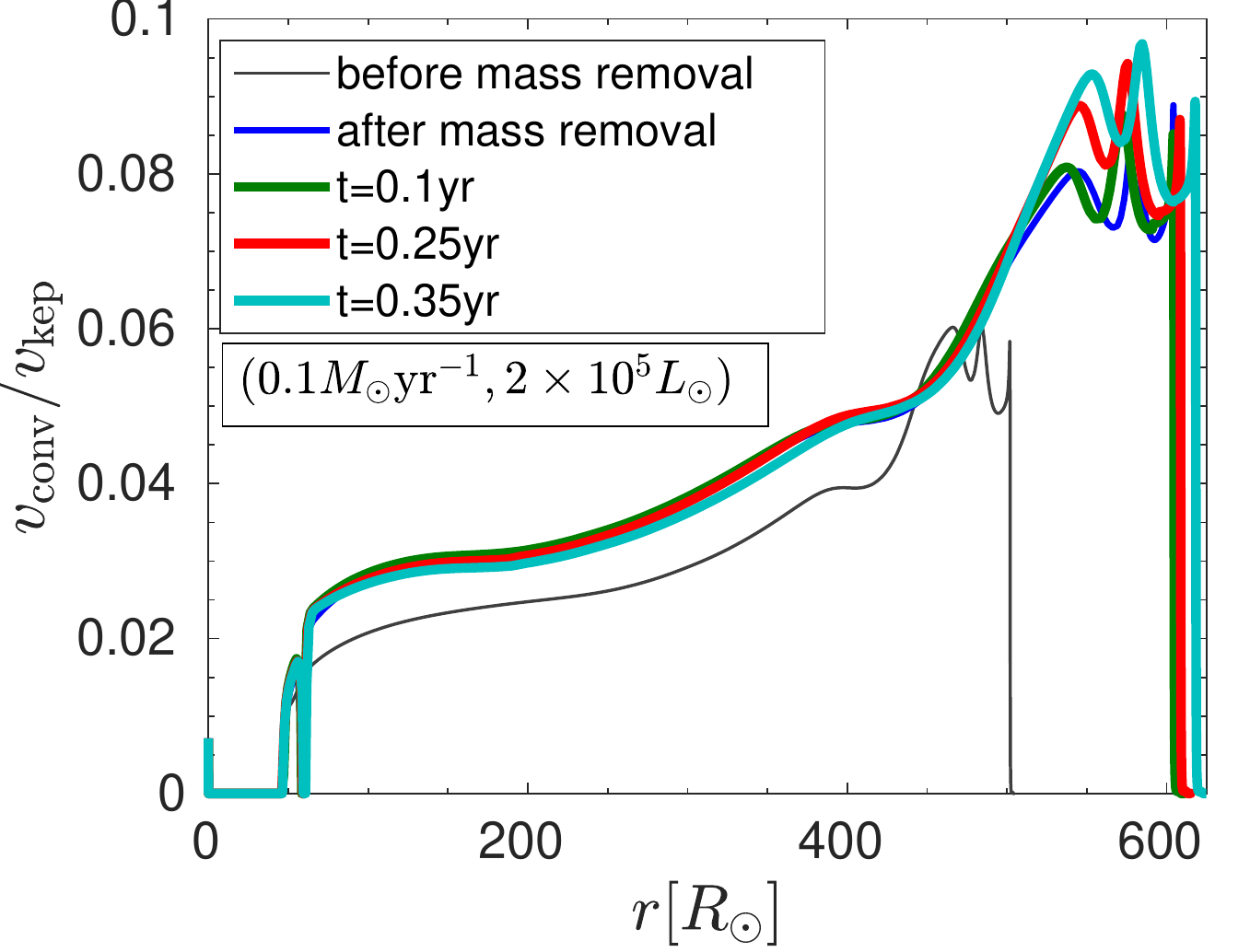}
\includegraphics[trim=3.2cm 9.4cm 4cm 9.9cm ,clip, scale=0.54]{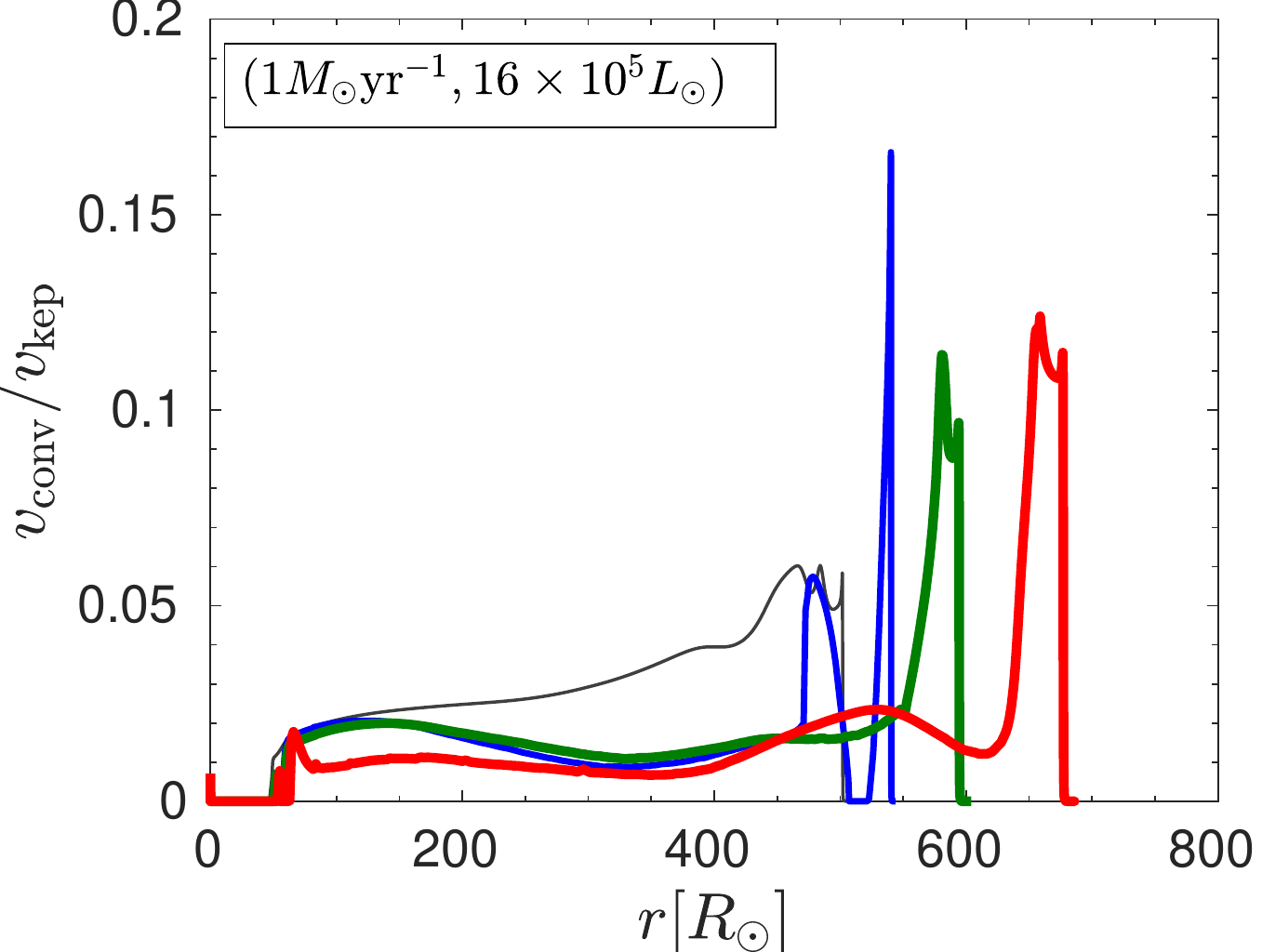}
\caption{Similar to Fig. \ref{fig:Vconv_93p_2mf}, but we stop removing mass and start to inject energy once $M_{\rm env} (0)=5.9 M_\odot$ (rather than $M_{\rm env} (0)=1.9 M_\odot$).}
\label{fig:Vconv_93p_6mf}
\end{figure}

In Fig. \ref{fig:Vconv_93p_2mf} we present $v_{\rm conv}(r)/v_{\rm Kep}(r)$ for a case where the envelope mass at  
the end of mass removal (which we set as $t=0$) is $M_{\rm env} = 
1.9 M_\odot$. The three different panels are for different mass removal rates and energy deposition powers, as indicated in the figure caption.  The thin-solid-black line in each panel presents the ratio just before we start the mass removal. 
The thick-blue line at each panel presents this ratio just at the end of mass removal ($t=0$). The thick colors lines present the ratio at different times after energy deposition at the indicated power and into a shell of $0.07 R_1(t) < r < R_1(t)$. In Fig. \ref{fig:Vconv_75p_2mf} we present the same ratio but for cases where the energy deposition is into a shell of $0.25 R_1(t) < r < R_1(t)$. In Fig. \ref{fig:Vconv_93p_6mf} we present this ratio for a cases that leave a larger envelope mass of $M_{\rm env} = 5.9 M_\odot$.   

\begin{figure}[t]
	\centering
\includegraphics[trim=3.2cm 9.4cm 4cm 9.9cm ,clip, scale=0.54]{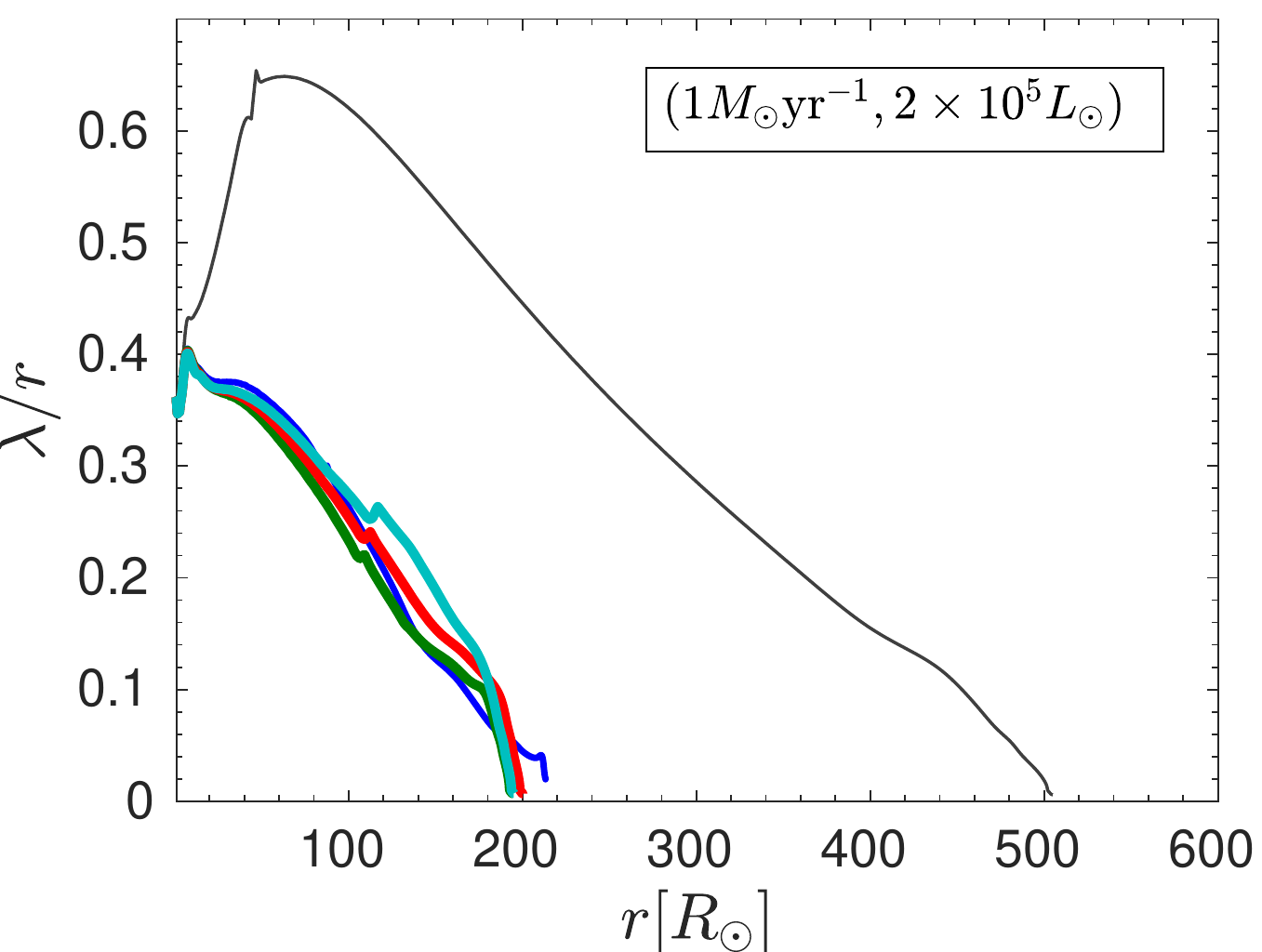}
\includegraphics[trim=3.2cm 9.3cm 4cm 9.9cm ,clip, scale=0.54]{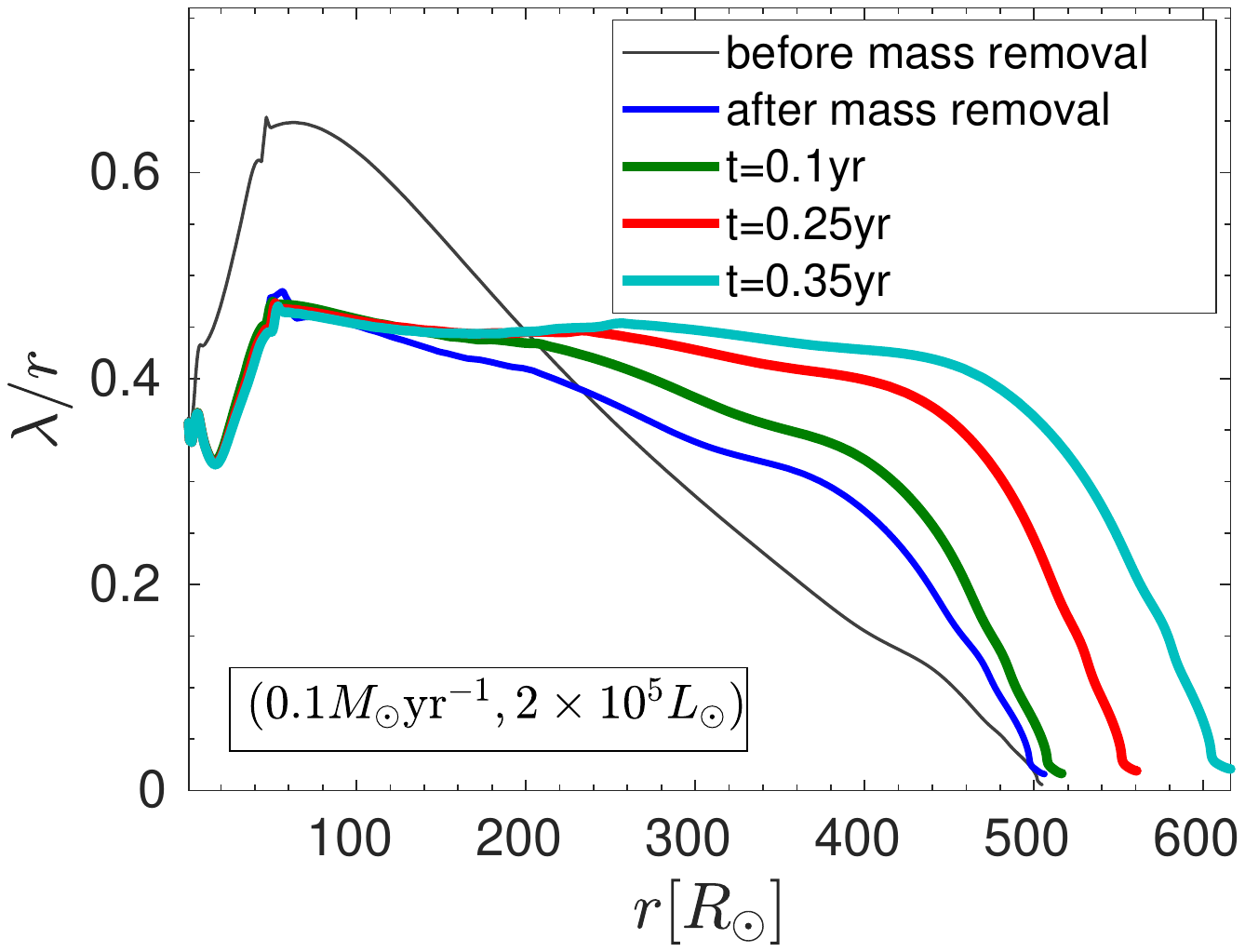}
\includegraphics[trim=3.2cm 9.4cm 4cm 9.9cm ,clip, scale=0.54]{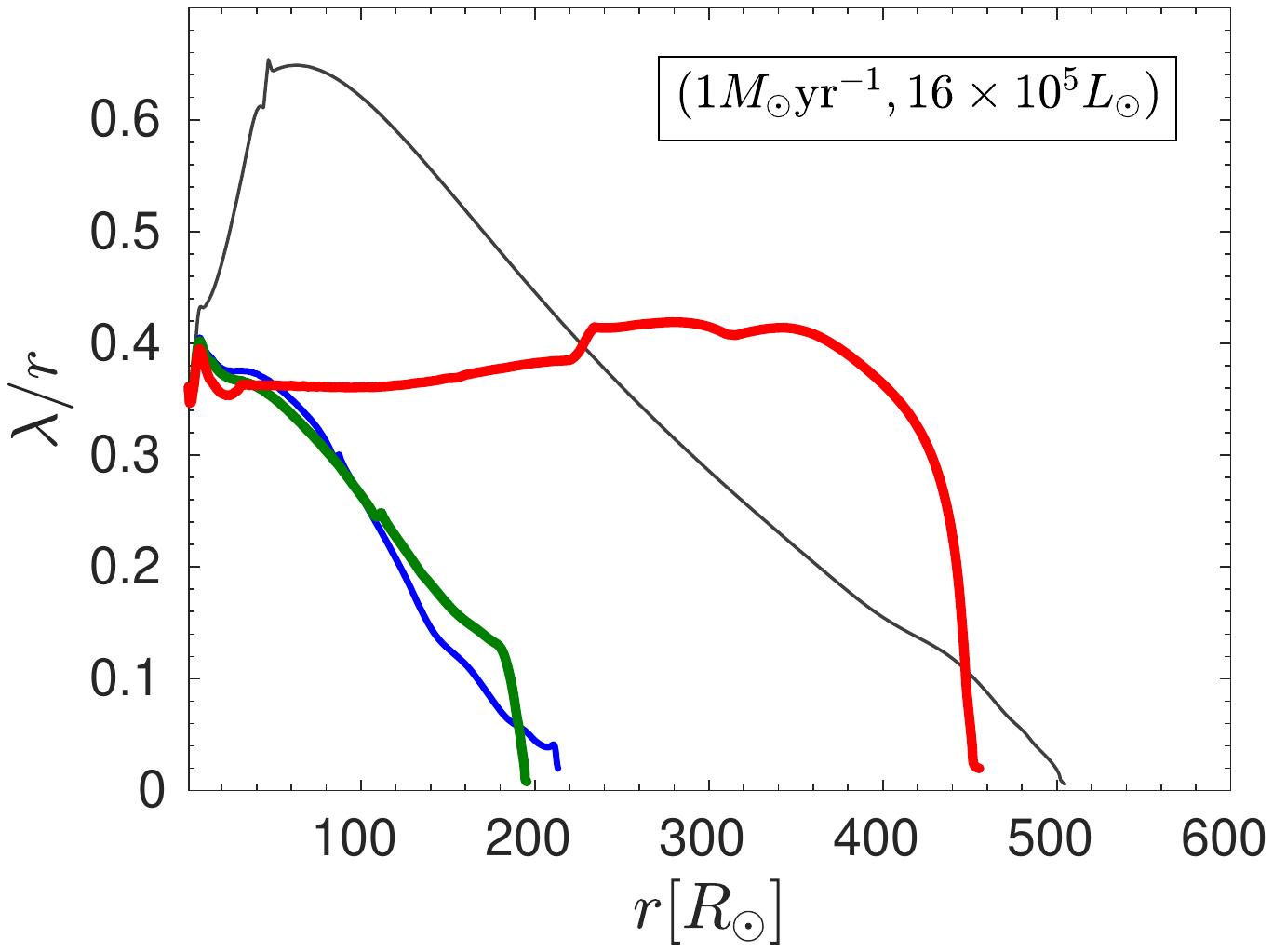}
\caption{The ratio of the mixing length to the radius, $\lambda(r)/r$, as a function of radius inside the envelope for the same cases and times as in Fig. \ref{fig:Vconv_93p_2mf}. }
\label{fig:Lmix_93p_2mf}
\end{figure}

In Fig. \ref{fig:Lmix_93p_2mf} we present the ratio of $\lambda/r$ for the same cases as in Fig. \ref{fig:Vconv_93p_2mf}.  
 
Figs. \ref{fig:Vconv_93p_2mf} - \ref{fig:Vconv_93p_6mf} show that before we start rapid mass removal the convective velocity in the outer envelope is $v_{\rm conv}(r)/v_{\rm Kep}(r) \simeq 0.02-0.05$ (note the different scales of the different panels).  
Rapid mass removal changes this ratio and the envelope radius, depending on mass removal rate $\dot M$. When this rate is low, $\dot M = 0.1M_\odot\yr^{-1}$ (as we present in the second panels of Figs. \ref{fig:Vconv_93p_2mf} - \ref{fig:Vconv_93p_6mf}), the ratio increases by a factor of up to 2-3, namely, $v_{\rm conv}(r)/v_{\rm Kep}(r) \simeq 0.02-0.15$. The higher increase are for even lower mass removal rates than we show here (below what we expect in a CEE with a NS/BH companion). However, at $\dot M=1M_\odot\yr^{-1}$ (upper panels in the three figures) the ratio decreases somewhat as the envelope shrinks. Clearly as we deposit energy the ratio $v_{\rm conv}(r)/v_{\rm Kep}(r)$ increases and the star drastically expands (green, red, cyan thick lines in Figs. \ref{fig:Vconv_93p_2mf} - \ref{fig:Vconv_93p_6mf}).

As expected, when the envelope mass is larger the same energy causes smaller changes, as can be seen in Fig. \ref{fig:Vconv_93p_6mf} that has a larger envelope mass, $M_{\rm env}=5.9M_\odot$ instead of $M_{\rm env}=1.9M_\odot$. 

In simulations that we do not show here we found that if we deposit energy at the same rates as in the simulations we present here but we do not remove mass from the envelope, then no significant changes occurs in the envelope structure and convective velocities.  This implies that envelope mass removal causes larger changes to the relevant convective properties than energy deposition in the CEE that we consider here.  

Fig. \ref{fig:Vconv_75p_2mf} presents simulations where we deposited the energy into lower envelope mass, i.e., in outer layers, as compared to the simulations of Fig. \ref{fig:Vconv_93p_2mf}. As expected, this results in a larger impact of energy deposition on the outer envelope, i.e., larger expansion of the envelope and much larger values of $v_{\rm conv}(r)/v_{\rm Kep}(r)$. 
 
The highest energy deposition rate we simulate here,  $L=16\times10^5L_\odot$ (lower panels) leads to evolution at late times that the numerical code cannot handle. This includes an expansion velocity larger than the escape speed and a convective velocity larger than the Keplerian velocity. For that we do not present the structure at $t=0.35 \yr$ in these cases. Such energy deposition rates require hydrodynamical simulations (see references in section \ref{sec:intro}). 

Fig. \ref{fig:Lmix_93p_2mf} (and others that we do not present) show that the mixing length in the different cases after mass removal and energy deposition suffers smaller changes than the convective velocity. 

Overall, our results that will serve us in section \ref{sec:CEE2} show that the typical scaling ratios relevant to our study are $v_{\rm conv}(r) \simeq 0.1 v_{\rm Kep}(r)$ and $\lambda(r) \simeq 0.3 r$. 

\section{Implications on mass accretion in CEE}
\label{sec:CEE2}

\begin{figure}[t]
	\centering
\includegraphics[trim=3cm 9.4cm 4cm 9.9cm ,clip, scale=0.62]{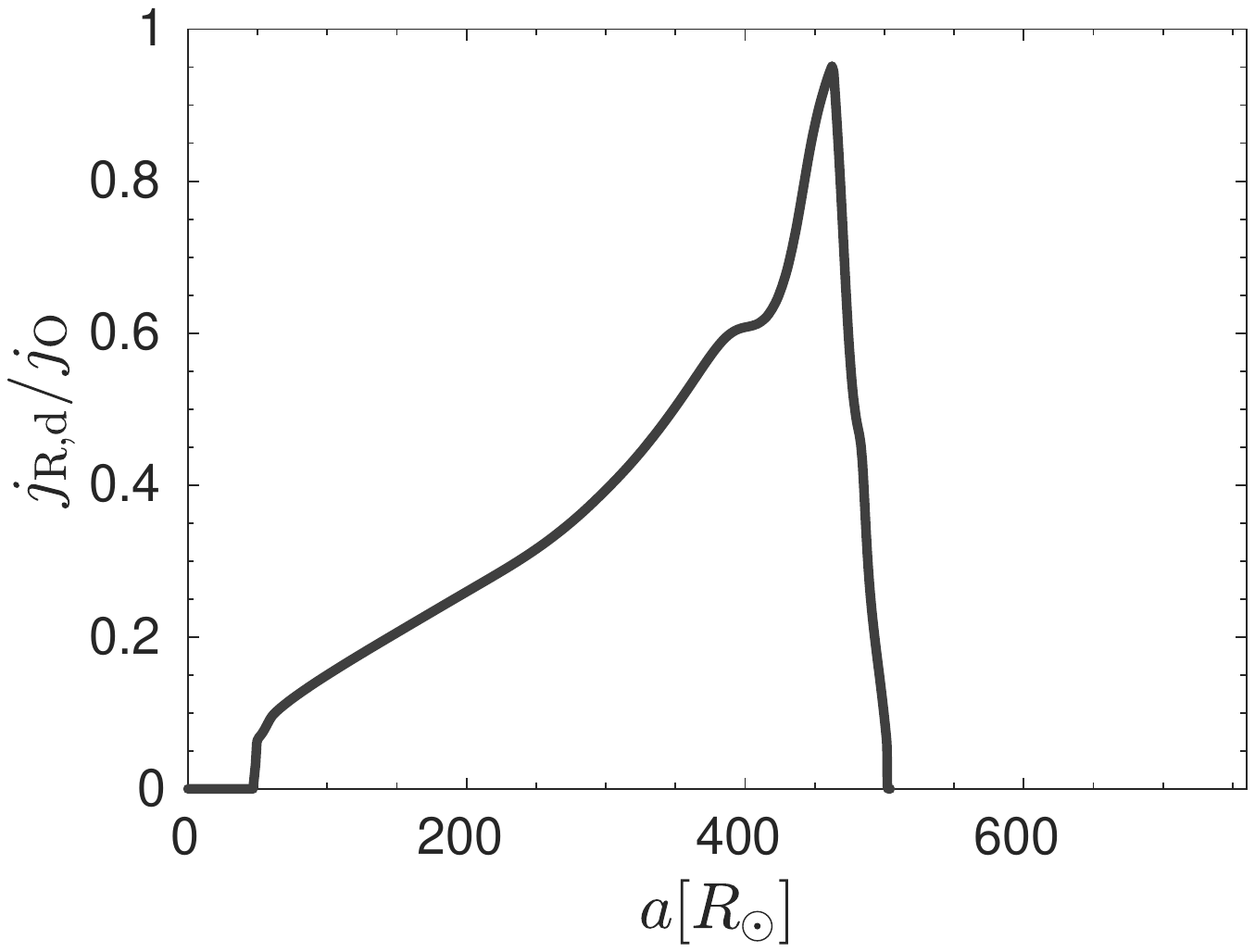}
\includegraphics[trim=3cm 9.4cm 4cm 9.9cm ,clip, scale=0.62]{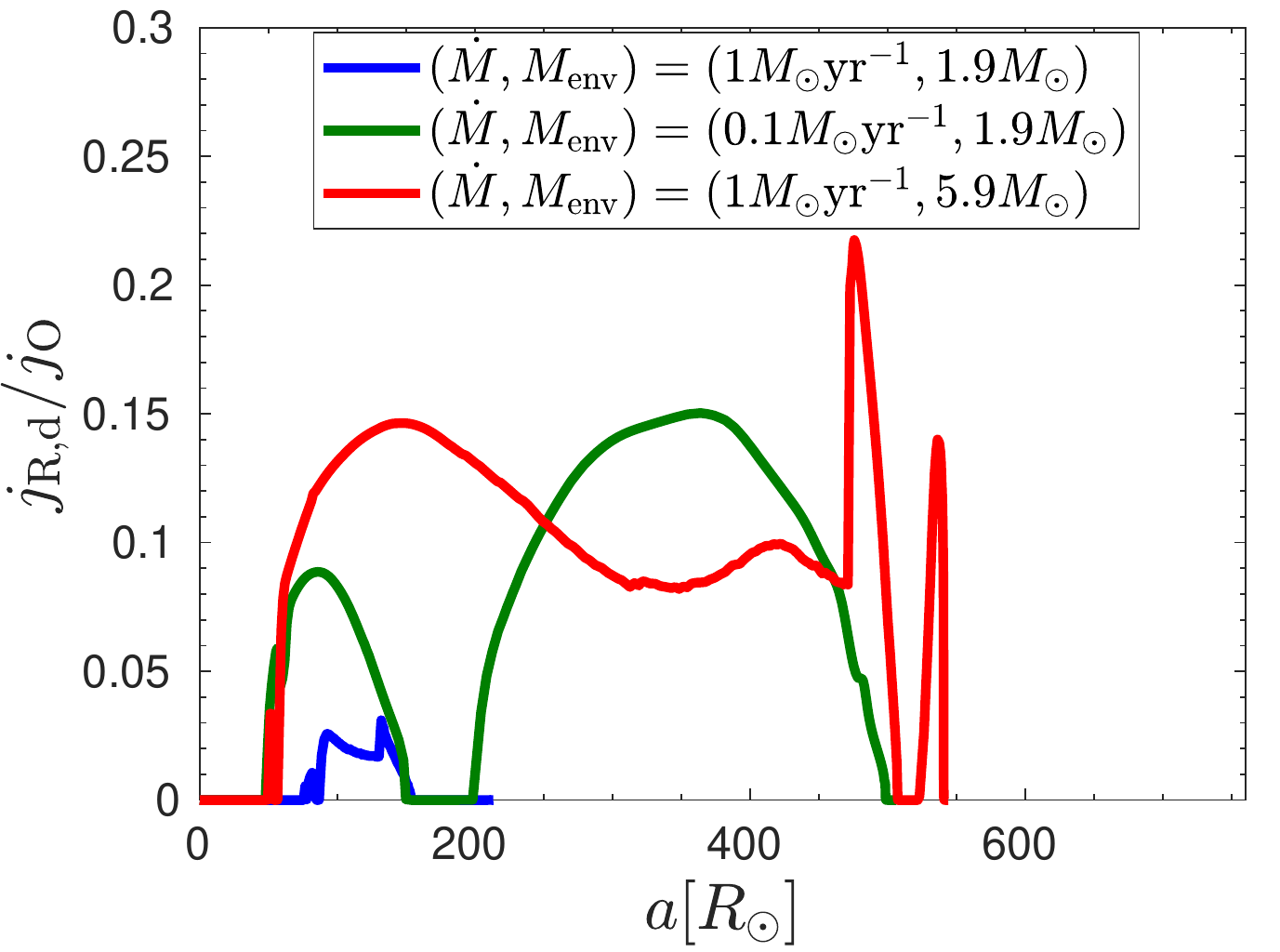}
\caption{The ratio of the stochastic specific angular momentum of the accreted mass to the one that results from the orbital motion and density gradient in the envelope, $j_{\rm R,d}/j_{\rm O}$, as a function of orbital separation, according to equation (\ref{eq:jRatioD}) and for a neutron star companion of mass $M_2=1.4 M_\odot$. Models are for a star with initial mass of $M_{\rm ZAMS}=15M_\odot$ that we evolved in section \ref{sec:Convective}. 
Upper panel: The ratio for the undisturbed star evolved up to $R=500R_\odot$. At this point $M_1=14.7M_\odot$ and the envelope mass is $M_{\rm env}=11.4 M_\odot$.  
Lower panel: The same star, but after we removed mass at a high rate of $\dot M$ to $M_{\rm env} < 11.4 M_\odot$, for three cases as indicated in the inset. 
}
\label{fig:Jrd_Jo_undisturbed}
\end{figure}

\begin{figure}[t]
	\centering
\includegraphics[trim=3cm 9.4cm 4cm 9.9cm ,clip, scale=0.62]{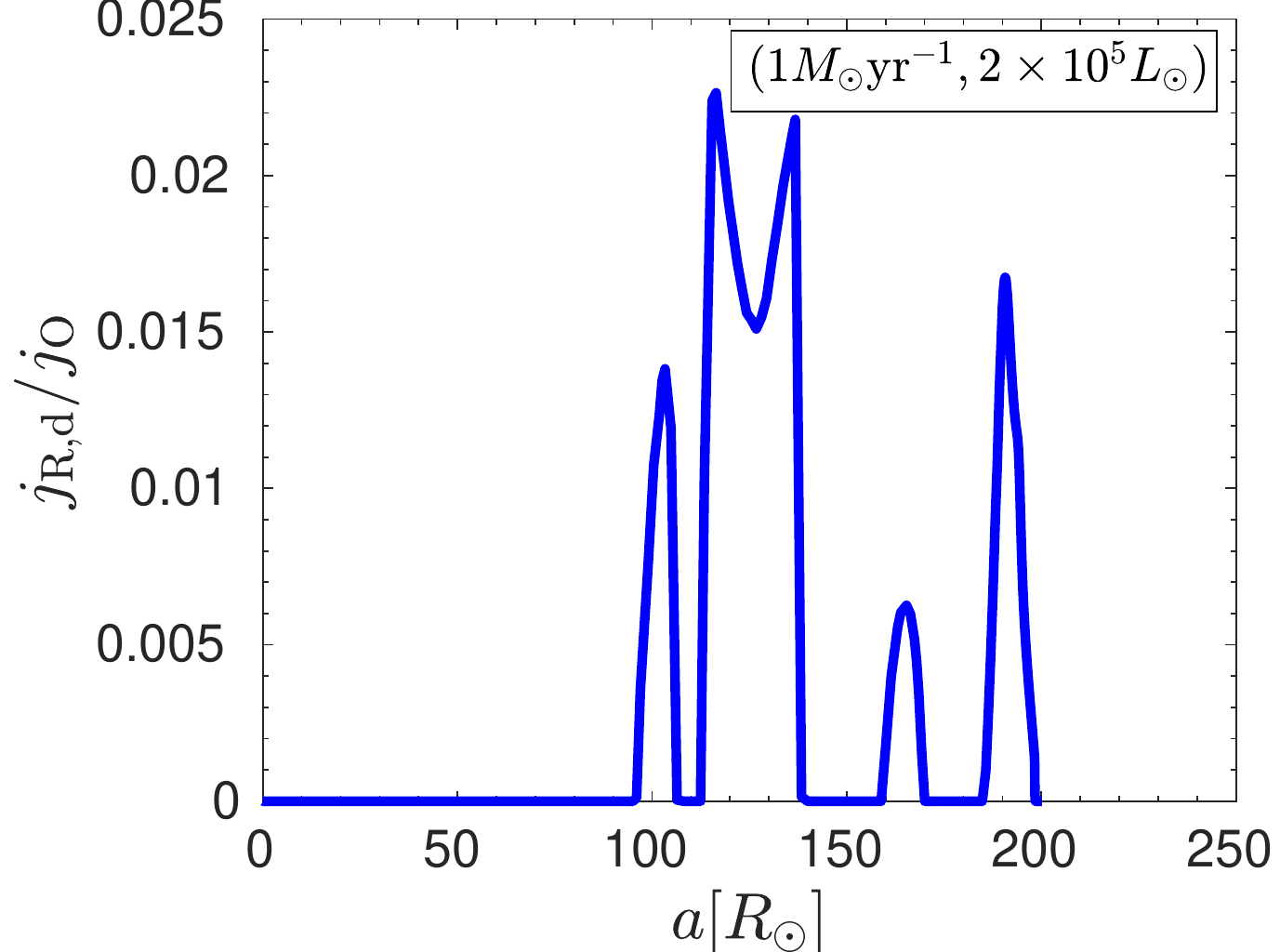}
\includegraphics[trim=3cm 9.4cm 4cm 9.9cm ,clip, scale=0.62]{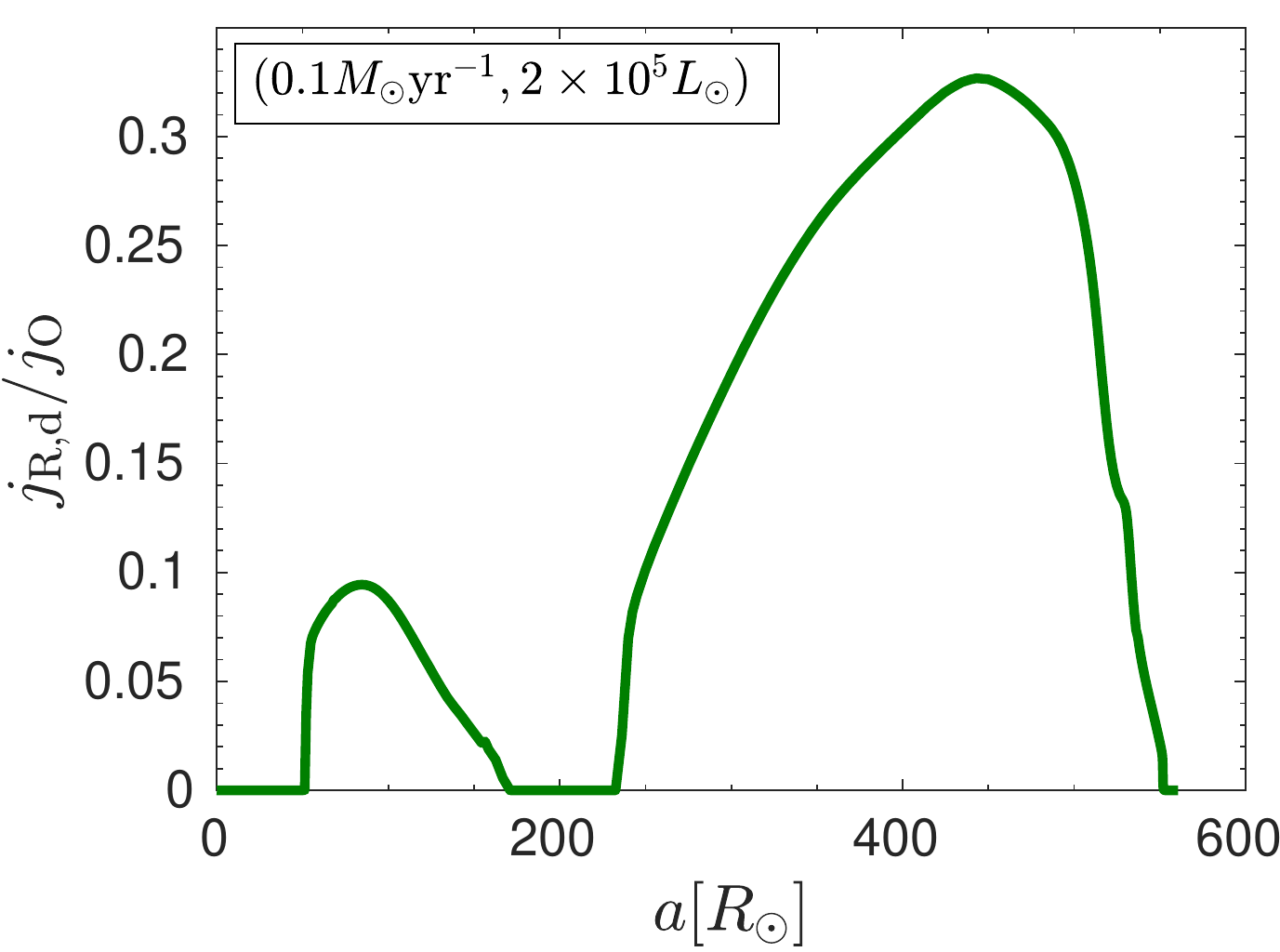} 
\includegraphics[trim=3cm 9.4cm 4cm 9.9cm ,clip, scale=0.62]{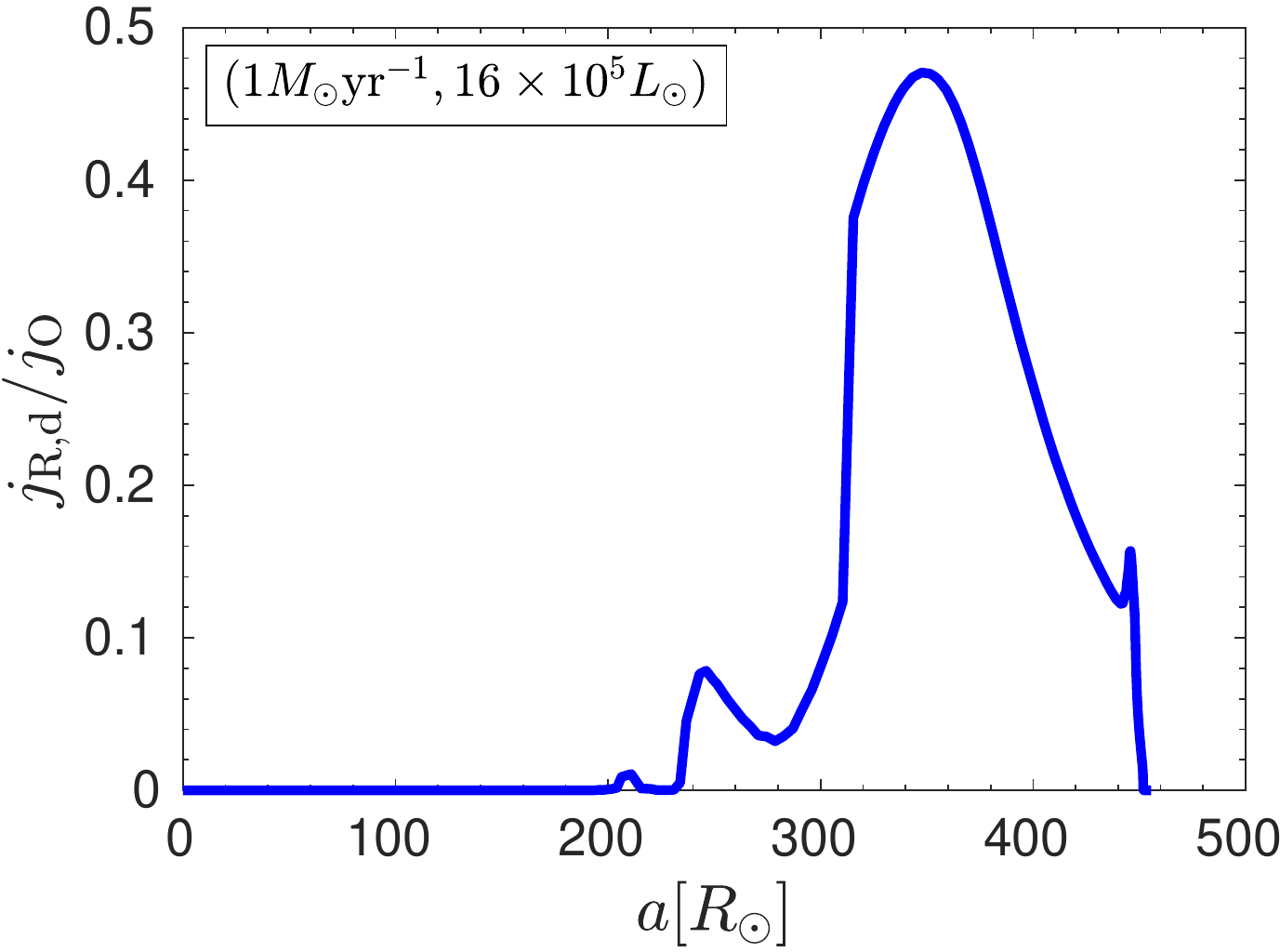}
\caption{Similar to Fig. \ref{fig:Jrd_Jo_undisturbed} but after we deposited energy to the envelope following mass removal at a rate of $\dot E$ and as in Figs. \ref{fig:Vconv_93p_2mf} and \ref{fig:Lmix_93p_2mf}. All panels are at $t=0.25 \yr$, i.e., we deposited energy for $0.25 \yr$. In all cases the envelope mass is $M_{\rm env}=1.9 M_\odot$. The cases are for $(\dot M, \dot E)$ values as given in the insets. }
\label{fig:Jrd_Jo_disturbed}
\end{figure}

In Fig. \ref{fig:Jrd_Jo_undisturbed} we present the ratio $j_{\rm R,d}/j_{\rm O}$ as given by equation (\ref{eq:jRatioD}), and where the different variables are taken for a secondary mass of $M_2=1.4 M_\odot$, which represents a NS, and for the RSG models that we simulated in section \ref{sec:Convective}.  
In the upper panel we take the values of $M_1(a)$, $\lambda$ and $v_{\rm conv}$ at each radius of the unperturbed stellar model when $M_1=14.7 M_\odot$ (before rapid mass removal and before energy injection). We keep the values of $\zeta$ and $\eta$ as in the scaled equation. For the density profile (equation \ref{eq:EnvDensty}) we take the average in the envelope which is $\beta=3.3$.
In the lower panel we show this ratio according to equation (\ref{eq:jRatioD}) and for $M_2=1.4 M_\odot$ for three cases after rapid mass removal, but before energy injection. The envelope mass removal rate and the final envelope mass are given in the inset. After mass removal we find  $\beta=3.2$.  Overall, the value of $\beta$ does not change much and the change in its value during the evolution has a very small influence on the results.  In Fig. \ref{fig:Jrd_Jo_disturbed} we show this ratio for three cases after we deposited energy to the envelope (see caption for the parameters). 
The convective velocities of the different models can be found in Figs. \ref{fig:Vconv_93p_2mf} to \ref{fig:Vconv_93p_6mf}, and the mixing lengths for one model in Fig. \ref{fig:Lmix_93p_2mf}. 

\begin{figure}[t]
	\centering
\includegraphics[trim=3cm 9.4cm 4cm 9.9cm ,clip, scale=0.62]{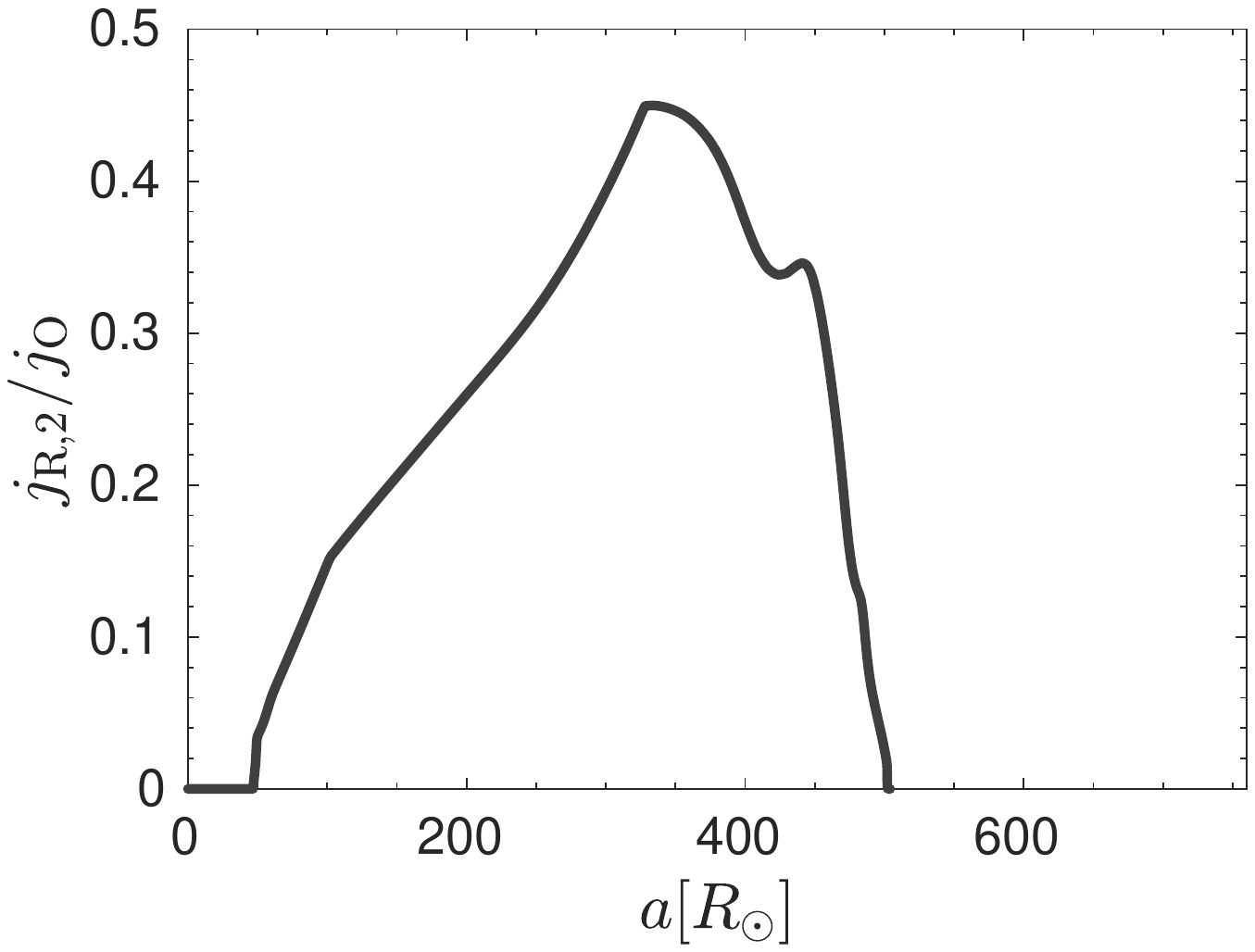}
\includegraphics[trim=3cm 9.4cm 4cm 9.9cm ,clip, scale=0.62]{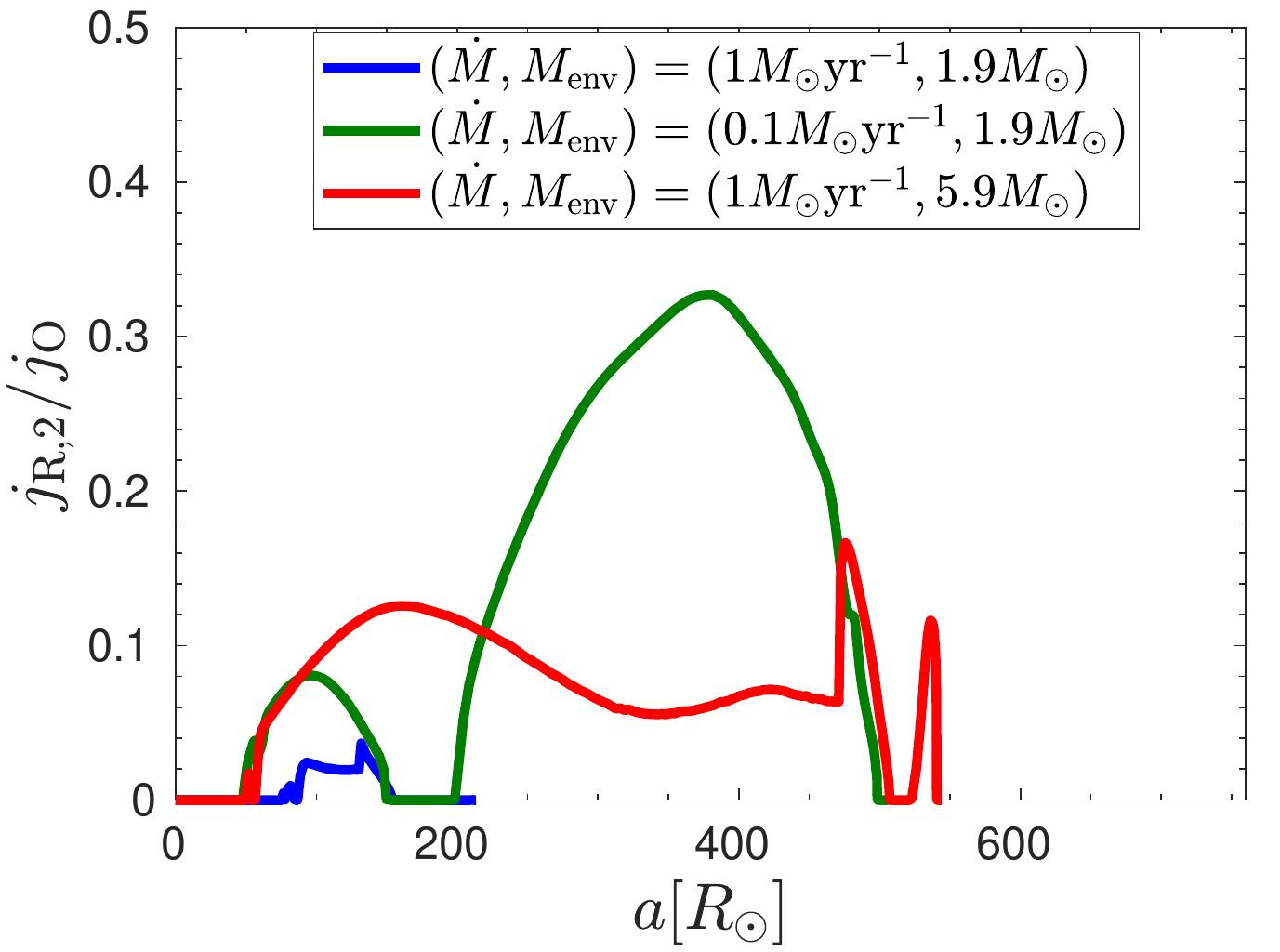}
\caption{ Similar to Fig. \ref{fig:Jrd_Jo_undisturbed} but for the ratio $j_{\rm R,2}/j_{\rm O}$ as given by equation (\ref{eq:jRatio2}) which is appropriate for a main sequence companion. We take $M_2=1.4 M_\odot$, but now representing a main sequence star. 
}
\label{fig:Jr2_Jo_undisturbed}
\end{figure}

\begin{figure}[t]
	\centering
\includegraphics[trim=3cm 9.4cm 4cm 9.9cm ,clip, scale=0.62]{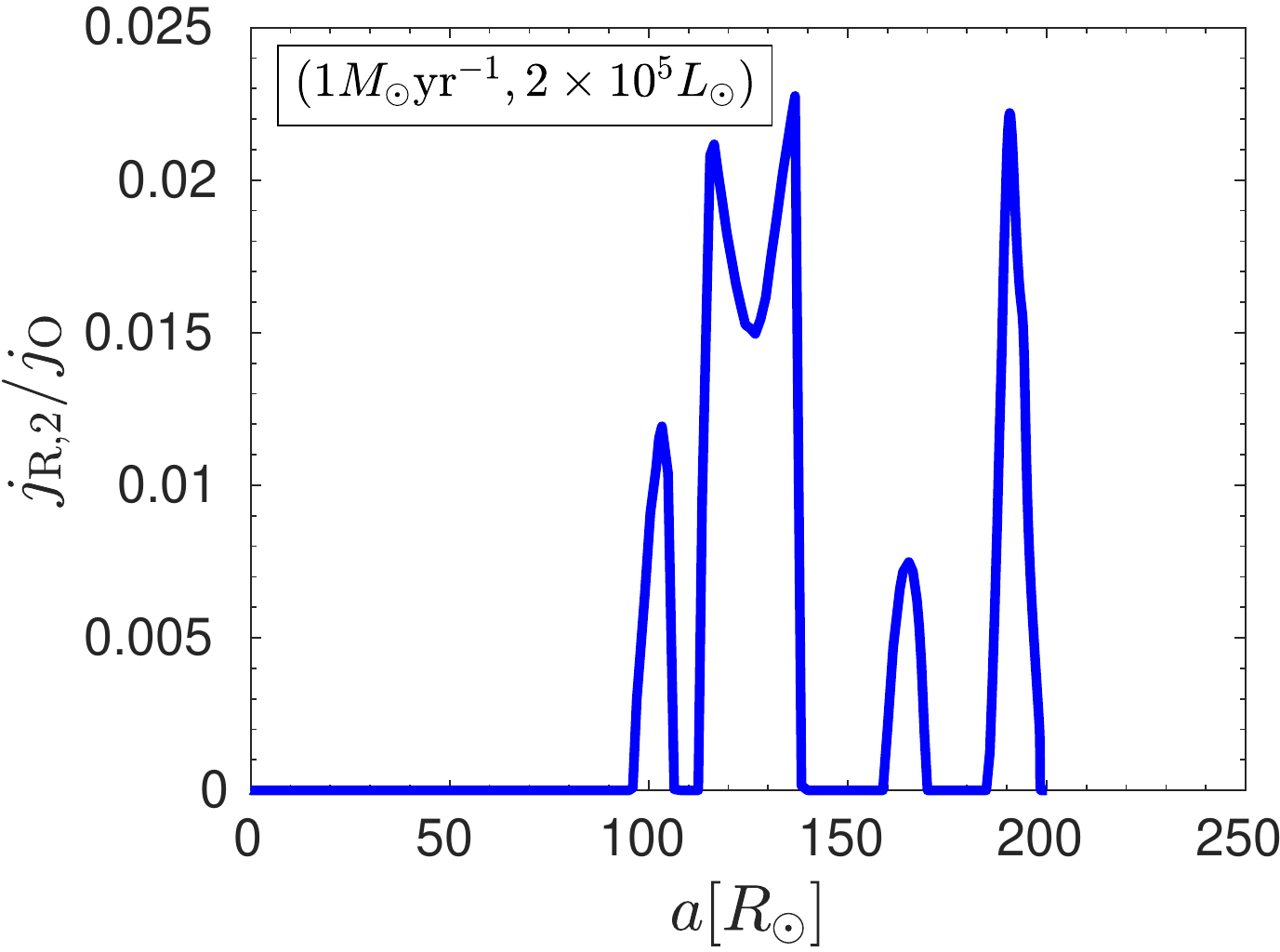}
\includegraphics[trim=3cm 9.4cm 4cm 9.9cm ,clip, scale=0.62]{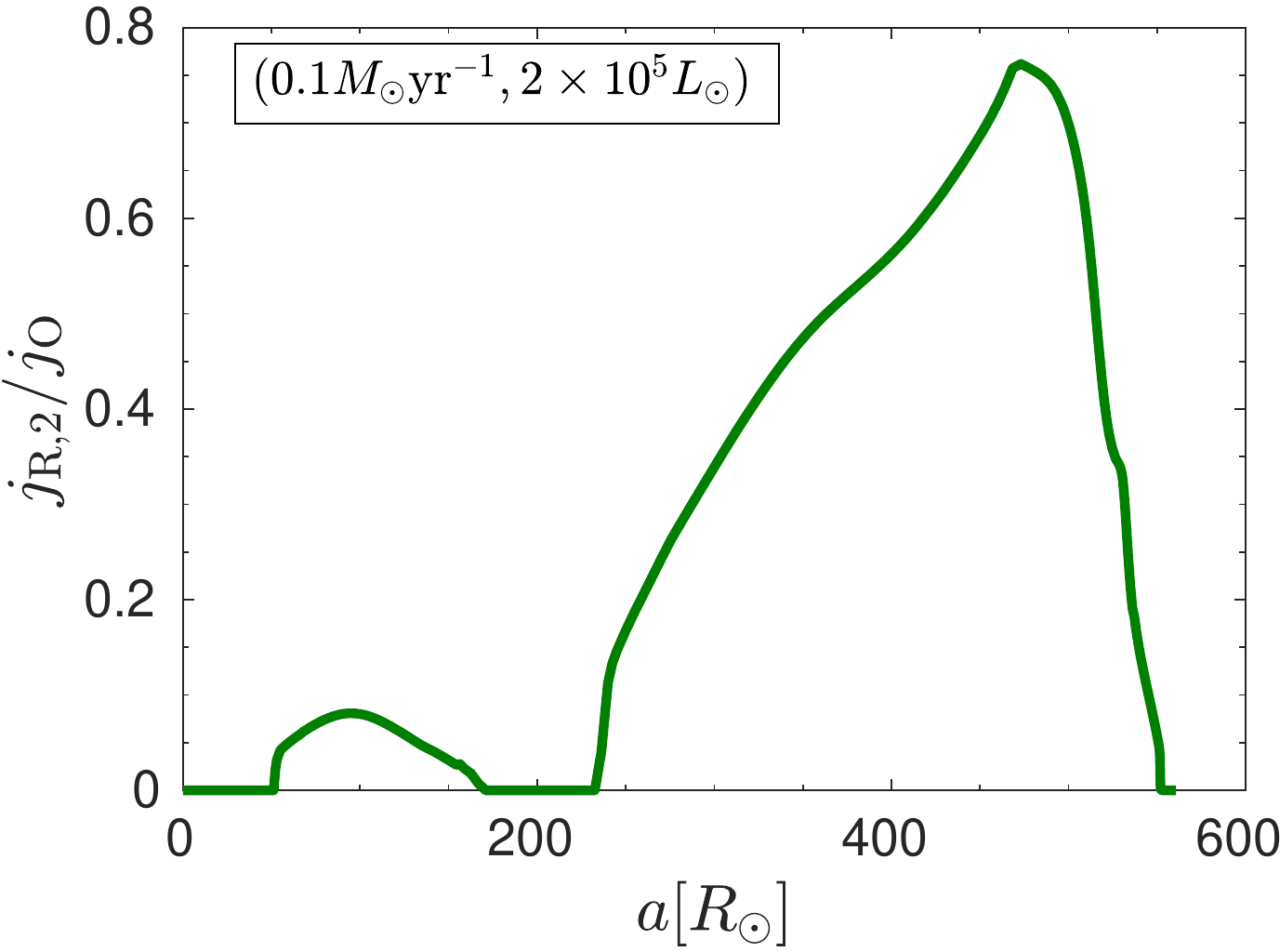} 
\includegraphics[trim=3cm 9.4cm 4cm 9.9cm ,clip, scale=0.62]{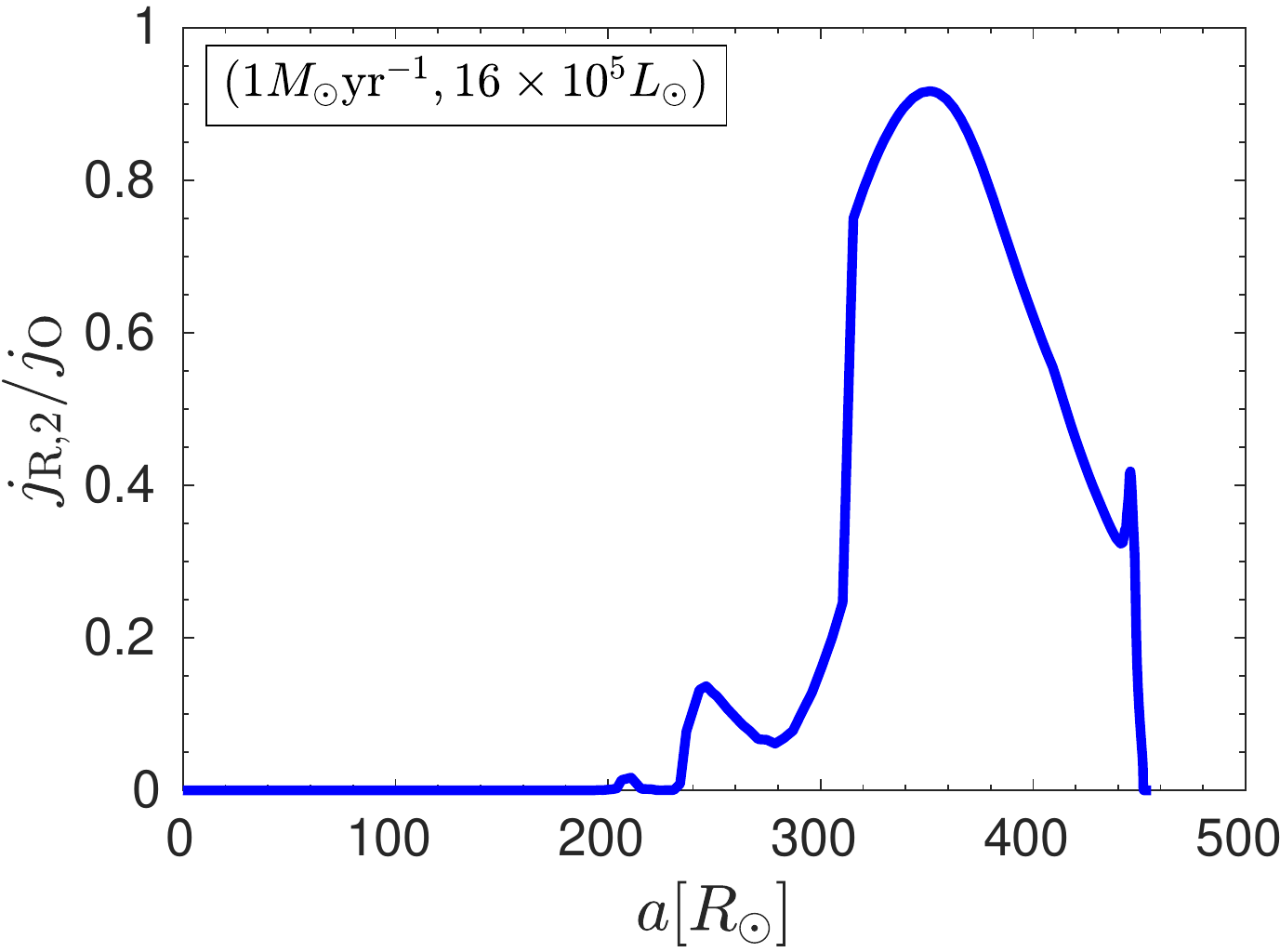}
\caption{ Similar to Fig. \ref{fig:Jrd_Jo_disturbed}
but for the ratio $j_{\rm R,2}/j_{\rm O}$ as given by equation (\ref{eq:jRatio2}) which is appropriate for a main sequence companion. We take $M_2=1.4 M_\odot$, but now representing a main sequence star as in Fig. \ref{fig:Jr2_Jo_undisturbed}.
}
\label{fig:Jr2_Jo_disturbed}
\end{figure}

We also calculate the ratio $j_{\rm R,2}/j_{\rm O}$ as given by equation (\ref{eq:jRatio2}) which is appropriate for a main sequence companion. We take $M_2=1.4 M_\odot$ as above, but now it is a main sequence star rather than a NS. 
In Fig. \ref{fig:Jr2_Jo_undisturbed} we present this ratio before energy deposition, in the upper panel before mass removal and in the lower panel after mass removal. 
In Fig. \ref{fig:Jr2_Jo_disturbed} we present this ratio after energy deposition. 

The envelopes of RGB, AGB, and RSG stars are known to have vigorous convection. In section \ref{sec:Convective} we determined the convective properties in RSG envelopes when we remove mass and add energy as expected in CEE. We used the typical convective properties that we found to scale equations (\ref{eq:jRatio2}) and (\ref{eq:jRatioD}). 
 With this appropriate scaling equations (\ref{eq:jRatio2}) and (\ref{eq:jRatioD})  show that the stochastic angular momentum of the accreted gas due to the convective motion might set the jets that the secondary star launches to stochastically change direction around the orbital angular momentum direction, which is perpendicular to the equatorial plane.

We further calculated the ratio of the random stochastic specific angular momentum component to the one perpendicular to the orbital plane due to the orbital motion as function of radius. Namely, we calculated the ratios in equations (\ref{eq:jRatio2}) and (\ref{eq:jRatioD}) as function of radius for the stellar models that we developed in section \ref{sec:Convective}. The ratio $j_{\rm R,2}/j_{\rm O}$ by equation (\ref{eq:jRatio2}) is appropriate for a main sequence companion while the ratio $j_{\rm R,d}/j_{\rm O}$ by equation (\ref{eq:jRatioD}) is appropriate for a NS companion. 
We present the results in Figs. \ref{fig:Jrd_Jo_undisturbed} - \ref{fig:Jr2_Jo_disturbed}. 

From these figures we learn the following. 
(1) The typical values for the parameters we use here are 
$j_{\rm R,2}/j_{\rm O} \simeq 0.1-1$ and $j_{\rm R,d}/j_{\rm O} \simeq 0.1-1$. The smaller values are for cases where the power of the jets is very low, lower than we expect for CEE with rapid mass removal. (2) The ratios depend on the mass removal rate, being larger for slower mass removal. 
(3) The ratio increase with increasing jets' power $\dot E$. 
(4) Large values of the ratios exist in the outer zones of the envelope. 

Overall, we expect the jets to wobble around the orbital angular momentum axis (perpendicular to the orbital planet). We discuss the implication of the wobbling jets in the next section.

\section{Discussion and Summary}
\label{sec:Summary}

Jets that the more compact companion launches during CEE as it accretes mass from the extended envelope of a giant star might play significant roles (section \ref{sec:intro}).  
The jets operate in a feedback mechanism to regulate the mass accretion rate, they can facilitate envelope removal, they influence the morphology of the ejecta,  and they can power a bright transient event. A main sequence companion can launch jets that power an intermediate luminosity optical transient (ILOT; also termed luminous red nova; e.g., \citealt{Soker2015GEE}), while a NS/BH companion can power a transient event that mimics a core collapse supernova (termed common envelope jets supernova; e.g., \citealt{SokerGilkis2018, GrichenerSoker2021}).  

The orbital motion of the companion inside the envelope has a large scale mirror-symmetry about the equatorial plane. This by itself would form an accretion disk (or an accretion belt) in the orbital plane, implying that the jets are launched perpendicular to the orbital plane, i.e., along the orbital angular momentum direction. 

In this study we found that the convective motion in the envelopes of RSG stars supplies a non-negligible stochastic angular momentum to the mass that the secondary star accretes. The magnitude of the stochastic specific angular momentum component is up to $\simeq 0.1-1$ times the specific angular momentum of the accreted mass due to the orbital motion, i.e., $j_{\rm R} \simeq (0.1-1) j_{\rm O}$ (section \ref{sec:CEE2}; Figs. \ref{fig:Jrd_Jo_undisturbed} - \ref{fig:Jr2_Jo_disturbed}). This large stochastic angular momentum variation leads to the wobbling of the accretion disk/belt and hence to the wobbling of the jets around the orbital angular momentum in angles of up to $\tan ^{-1}({j_{\rm R,2}}/{j_{\rm O}}) \simeq 10^\circ - 40 ^\circ$ or 
$\tan ^{-1}({j_{\rm R,d}}/{j_{\rm O}}) \simeq 10^\circ - 40 ^\circ$. 

 We expect that in the case of RGB and AGB stars the results of the present study regarding the wobbling of the jets will be similar, because RSG and AGB stars also have strong envelope convection. The main difference is that many RSGs engulf NSs, BH, or massive main sequence stars (that have radiative envelope), while RGB and AGB stars likely engulf white dwarfs or low-mass main sequence stars (that have convective envelope). The power of the jets will be much higher in the case of RSGs that engulf NSs or BHs. This is the reason we concentrate on RSG stars in this study.  

The outcomes, like shaping of the ejecta and the amount of energy that the jets deposit to the envelope, strongly depend on whether the jets manage to break out from the envelope or not. The orbital motion of the compact object that launches the jets implies that the jets do not drill at a constant location. This in turn prevents the jets from breaking out of the envelope in many cases (e.g., \citealt{Papishetal2015, LopezCamaraetal2022MS}).
Our finding of large-amplitude wobbling implies that breaking out from the envelope (namely, penetrating the envelope) is less likely even. This makes energy deposition to the envelope more efficient, in particular when the companion is in the outer regions of the envelope. 

Large-amplitude wobbling influences the morphology of the ejecta. Jets with a constant-direction axis can form a bipolar nebula, i.e., inflate bubbles along a fixed symmetry axis. Jets that have large-amplitude wobbling, instead, might inflate small bubbles in the ejecta. The bubbles compress shells on their outskirts. These shells will form filaments and arcs in the descendant nebula. We therefore suggest that some arcs and filaments in planetary nebulae might result from wobbling jets (although we did not simulate AGB stars). 

Jets are more likely to inflate large lobes when the companion is outside the envelope at the time when it is entering the CEE. At that phase accretion rate is high, so the jets are powerful, and the jets might break out from the low-density envelope and shape the ejecta. It is possible that the convective motion in the outer envelope also leads to wobbling jets at that phase. Three-dimensional hydrodynamical simulations are required to explore the shaping by wobbling jets. 

Overall, our study adds to the complexity and rich variety of process that take place in CEE with jets. Specifically, the wobbling of the jets that the compact companion in a CEE launches implies that the jets deposit more of their energy in the envelope and shape small bubbles and filaments in the ejecta. The jets in a CEE are less likely to maintain a constant axis. This might have  implications to the time variability of very-high-energy (PeV range) neutrinos in the common envelope jets supernova scenario for such neutrinos where the compact companion is a black hole (\citealt{GrichenerSoker2021, Grichener2023}).


\section*{Acknowledgments}

 We thank an anonymous referee for helpful comments.  This research was supported by a grant from the Israel Science Foundation (769/20).



\label{lastpage}

\end{document}